\documentclass[draft]{agujournal2019}
\usepackage{url} 
\usepackage{lineno}
\usepackage{caption}
\usepackage[inline]{trackchanges} 
\usepackage{soul}
\usepackage{amsmath, amssymb}
\usepackage{lmodern}
\usepackage{float}

\draftfalse

\journalname{Journal of Advances in Modeling Earth Systems (JAMES)}

\begin{document}

\title{A Self-Diagnosing Structural Error-Aware Parameter Estimation Method for Earth System Models}

\authors{Qingyuan Yang\affil{1,2}, 
        Addisu G Semie \affil{1,2,3},
        Brian Medeiros \affil{1,3},
        Gregory S Elsaesser\affil{1,4,5},
        Da Fan \affil{1,2,3},
        Wayne Chuang \affil{1,2}}

\affiliation{1}{Learning the Earth with Artificial Intelligence and Physics (LEAP) National Science Foundation (NSF) Science and Technology Center, Columbia University, New York, NY, USA}
\affiliation{2}{Department of Earth and Environmental Engineering, Columbia University, New York, NY, USA}
\affiliation{3}{NSF National Center for Atmospheric Research, Boulder, CO, USA}
\affiliation{4}{NASA Goddard Institute for Space Studies, New York, NY, USA}
\affiliation{5}{Department of Applied Physics and Applied Mathematics, Columbia University, New York, NY, USA}

\correspondingauthor{Qingyuan Yang}{qy2288@columbia.edu}

\begin{keypoints}
\item We develop an efficient Earth System Model (ESM) autocalibration method that identifies and excludes the most structurally biased variables.  
\item The method is automated and enables model developers to learn how tolerating structural error may impact autocalibration. 
\item The method outputs diagnostics that enhance interpretability and reveal how parameters are constrained. 

\end{keypoints}

\begin{abstract}
We propose a fully automated, structural error-aware, interpretable climate model parameter estimation method that leverages Perturbed Parameter Ensembles (PPEs). It is based on history matching and aligns with an increasingly-used iterative simulation-emulation-calibration methodology. The method is motivated by the negative impacts of structural error and emulator and observational uncertainties on climate model parameter estimation efforts, as well as the problems associated with sparsely-sampled PPEs. To address these challenges, the method explicitly builds simpler emulators that avoid overfitting, detect structural error, avoids compensating for structural error through inflated mismatch tolerances, and sequentially excludes structurally inconsistent variables for parameter estimation. The method decomposes the high-dimensional calibration problem into linked low-dimensional subproblems, and integrates their constraints to reconstruct the jointly plausible region of the full parameter space. 
The method is applied to a 100-member PPE with 34 perturbed parameters generated by a version of CAM6 with machine learning-based warm rain microphysics parameterization. Through iterative application, the method greatly reduces the ensemble spread and improves the matching between simulated and observed zonal climatologies. The method also finds ensemble members that outperform the default CAM6 configuration in root mean square error across multiple diagnostics. Controlled experiments demonstrate that overly-conservative emulator uncertainty could lead to neglect of informative observations, and tolerance of the structural error, in the context of this method, biases the estimated parameters toward compensating for structural error. Our work also emphasizes the value of interpretability for diagnosing structural error and informing parameter estimation in PPE-based calibration.
\end{abstract}

\section*{Plain Language Summary}
Climate models use adjustable numbers (``parameters'') to simplify and represent certain physical processes (e.g., cloud formation). We run a climate model many times with different parameter values and compare the outputs to observations to find the best-fitting parameter values. Here, one of the challenges is that the model could have internal structural errors that cannot be addressed by varying the parameter values. This structural error can lead to compensating effects, for example, improving how the model output matches the observed temperature while degrading its simulation of precipitation. In this work, we design a method that detects which variables bear the largest structural error, and excludes them from the parameter estimation effort. The method decomposes a 34-parameter search into multiple two-parameter searches and reassembles them. The method is applied to an atmosphere model. It narrowed plausible parameter ranges substantially and matched satellite observations better than the model's defaults, while showing which variables drove each parameter value.

\section{Introduction}
A climate model Perturbed Parameter Ensemble (PPE) is a set of simulations generated by the climate model with varied model parameters \cite{rougier2009analyzing, sexton2012multivariate, williamson2013history,sexton2021perturbed,eidhammer2024extensible,bonnet2024tuning}. Each ensemble member uses the same model core but with a unique set of physics parameter values (e.g., autoconversion size threshold for atmospheric ice-snow transition) to generate a unique set of outputs (e.g., winds, temperature, and humidity, in the case of atmosphere-focused PPEs). These parameters could have a significant impact on the climate model outputs across various time scales, and their values are highly uncertain as a result of simplified representations of complex, unresolved physical processes \cite{graves1993new,neelin2010considerations,yang2012some}.

PPEs can be used to study the relationship between model parameters and climate model outputs \cite{regayre2015climatic,qian2018parametric, johnson2020robust,peatier2022investigating,peatier2024exploration}. They are used to study how the parameters affect global and regional climatologies such as cloud forcing and precipitation \cite{eidhammer2024extensible,yang2025simple}, identify model structural errors \cite{johnson2020robust, prevost2025detection,rostron2025clearer}, explore physical process hypotheses \cite{eidhammer2024extensible,gettelman2024interaction}, and characterize projection uncertainty \cite{peace2020effect,yamazaki2021perturbed,peace2022evaluating,duffy2024perturbing,watson2025integrating}.

PPEs are widely used for model calibration (i.e.,  parameter estimation) to select the parameter values that lead to the best model skill \cite{williamson2013history,karmalkar2019finding,sexton2021perturbed,dunbar2021calibration,carzon2023statistical,elsaesser2025using,bonnet2024tuning}. Compared to more traditional, manual model  calibration efforts involving changes in only a few parameter values at a time, numerous parameters within the PPEs are perturbed simultaneously across the members of the ensemble, enabling exploration of a large range of plausible model behaviors \cite{hourdin2017art,carslaw2025opinion}.

Due to the high computational cost of climate model simulations, the number of ensemble members (samples) within a PPE is typically limited to no more than a few hundred \cite{dunbar2022ensemble,yang2025simple}. This sampling density is sparse given that up to hundreds of parameters can be perturbed in a PPE or similar ensemble-based datasets (e.g., One-At-A-Time; \citeA{kennedy2025one}). The PPEs therefore in many cases cannot be \textit{directly} used for parameter estimation as any one member likely exhibits substantial deficiency in one or more metrics. They can, however, provide training data for emulators that act as surrogate models \cite{kennedy2001bayesian,rougier2009analyzing,bellprat2012objective,watson2021model}. The surrogate models, if well trained, can estimate the climate model outputs for a given set of parameter values at a negligible computational cost, facilitating efficient quantification of uncertainty. Numerous methods have been applied to PPEs to train surrogate models, including Neural Networks \cite{watson2021model,dagon2022machine,eidhammer2024extensible,elsaesser2025using}, Gaussian Processes (GP; \citeA{rasmussen2003gaussian,williamson2013history,bonnet2024tuning,johnson2020robust,yang2025simple}), and regression-based methods \cite{duffy2024perturbing,larson2025quadtune}.

With the emulators trained, parameter estimation can be formulated as an inverse problem in which the trained surrogate models (i.e., the emulators) can serve as the forward model \cite{tarantola2005inverse}. Within this framework, history matching and Bayesian approaches are widely applied to constrain the parameters based on observational data (which serve as the targets for parameter estimation) \cite{williamson2013history,bonnet2024tuning,elsaesser2025using}. The cycle of simulation-emulation-calibration \cite{williamson2013history,williamson2015identifying,elsaesser2025using} can be repeated multiple times, with the expectation that the parameters are progressively constrained toward the optimum region in the parameter space. Alternatively, non-emulator-based methods such as Ensemble Kalman Inversion and Gauss–Newton line-search algorithms directly iterate on the climate model for calibration without using surrogate models \cite{tett2017calibrating,dunbar2022ensemble,nicklas2025efficient}.

In practice, tightly constraining model parameters remains challenging in PPE-based calibration frameworks. The difficulty arises from multiple factors, including model structural error \cite{mcneall2016impact}, emulator uncertainty \cite{tebaldi2025emulators}, and observational uncertainty \cite{bellprat2017uncertainty,elsaesser2025using}. The structural error arises from missing processes, simplified dynamics, and imperfect functional forms adopted in physical parameterizations in climate models \cite{mcneall2016impact,eyring2024pushing}. Structural errors can be viewed as systematic model limitations (i.e., mismatches between model outputs and corresponding observations) that cannot be eliminated through varying the parameter values alone \cite{kennedy2001bayesian}. When a single target is considered, structural error may appear as the observation lying outside the ensemble-defined range. With multiple targets, it could also manifest as irreducible trade-offs, where no parameter set can simultaneously match all diagnostics. In practice, structural error is often difficult to disentangle from emulator uncertainty or observational product bias. Emulator uncertainty can sometimes partially mask the presence of structural error, while in other cases the magnitude of structural error could exceed the emulator’s predictive uncertainty. Observational uncertainty primarily reflects measurement noise, sampling limitations, and systematic retrieval errors, and is independent of the model parameterization itself \cite{rougier2009analyzing}, but for some cloud variables especially, biases could be mistaken for model structural errors \cite{elsaesser2025using}.

Structural errors have been explored through PPEs in \citeA{johnson2020robust,regayre2025remaining,elsaesser2025using,prevost2025detection,rostron2025clearer}, all of which collectively suggest that prioritizing certain target climatologies would lead to degraded model skills for others. These studies aim to reduce parametric uncertainty and projection spread while diagnosing the occurrence of structural errors. In the presence of unavoidable structural error, a central challenge is how to design effective and informative parameter estimation workflows. Numerous design choices are available at different stages of the workflow based on previous studies, including whether to use globally averaged climatologies or spatially resolved diagnostics as calibration targets \cite{watson2021model,elsaesser2025using,eidhammer2024extensible}, the choice of emulator design \cite{bellprat2013parameter,williamson2013history,yang2025simple}, and the parameter estimation method \cite{kennedy2001bayesian,williamson2013history}.  Existing calibration approaches for climate model PPEs typically accommodate structural error by inflating uncertainty tolerances or by seeking compromises across competing targets. While practical, these strategies can obscure the presence of structural inconsistencies and bias parameter estimates toward compensating for model deficiencies.

In this paper, we develop a new auto-calibration workflow that accounts for some of the issues discussed above. First, we introduce a structural error-aware auto-calibration strategy that detects and excludes inconsistent targets rather than absorbing their effects through uncertainty inflation. The process of excluding variables with structural error does not necessarily imply a loss of information, as spatially correlated and neighboring diagnostics without structural error can still provide indirect constraints on the parameters.  Second, we propose a low-dimensional decomposition approach that enables efficient parameter estimation under sparse sampling in high-dimensional parameter spaces. Third, we provide an interpretable workflow that explicitly links parameter constraints to individual diagnostics, facilitating diagnosis of structural error and calibration trade-offs.

Our workflow is conceptually related to \citeA{prevost2025detection}, who use a PPE to detect structural inconsistencies between observational constraints in an aerosol model. Their approach is diagnostic: it identifies conflicting constraints to inform model development priorities, but explicitly stops short of selecting a constrained parameter set or reducing parametric uncertainty. Our method instead treats structural error detection as an integral step within an automated, iterative calibration procedure that actively excludes inconsistent targets to constrain a usable parameter set. This distinction motivates four further design choices: (1) We do not spatially aggregating grid boxes into cluster means before comparison. We retain target variables at the individual latitude-band level, as increased aggregation can further obscure structural error; (2) We do not emulate the full parameter space jointly. We decompose calibration into low-dimensional subproblems that are recombined to reconstruct the plausible region in the full parameter space; (3) Instead of a fixed-percentile retention rule applied to the emulated ensemble, we adopt an uncertainty-scaled implausibility criterion that ties plausibility explicitly to the magnitude of emulator and observational uncertainty rather than a preset sample fraction; (4) Rather than working with a fixed number of emulated samples and resolving an empty intersection by relaxing the constraints, we adopt an adaptive sampling strategy that redraws candidate samples at substantially increased density when no plausible parameter sets are found, allowing an empty plausible region caused by insufficient sampling density to be distinguished from one caused by structural error. This method can be incorporated into the calibrating, emulating and sampling or calibrated parameter ensemble (CPE) model calibration frameworks \cite{cleary2021calibrate,elsaesser2025using}, and applies to both the initial PPE or the CPE. These iterative frameworks involve two levels of sampling: the sampled parameters used to run the climate model and the sampled parameters used to approximate the target observations, which is based on the emulator prediction. Throughout this work,``sampling'' or ``samples'' refers to the latter unless otherwise specified.

In the following text, we introduce our design choices, their justifications, and the method. We test our method on the estimation of 34 parameters in a version of CAM6 with Machine Learning-based warm rain microphysics parameterization (CAM6-ML) and its associated 100-member PPE. We analyze the performance of the method by examining the zonal climatologies and the global biases of the resultant calibrated ensemble members (with parameters selected by the method). We highlight the impacts of tolerating the structural errors and assuming overly conservative emulator uncertainty when the method is used, and discuss the limitations of our method.

\section{Method propositions and justifications}
We propose design considerations that account for sparsity, structural error, and observational and emulation uncertainty for parameter estimation utilizing climate model PPEs. We highlight the potential challenges that arise from adopting these choices.

\subsection{Parameter estimation targets}
Focusing on global mean quantities poses a problem in situations where regional opposite-sign biases cancel to yield a reasonable global mean, and will naturally bias calibration toward processes that dominate the Earth’s surface area (e.g., oceans) while partially neglecting the influence of parameters that affect localized processes. As a result, parameters with strong regional sensitivities may appear weakly constrained or unconstrained if global diagnostics are used as the parameter estimation targets. Targeting regional variables alongside globally averaged quantities has been adopted in previous works \cite{elsaesser2025using,larson2025quadtune,prevost2025detection}.

Increasing the number of calibration targets (e.g., the inclusion of more regional climatologies) may reveal structural errors that would otherwise remain obscured by averaged or more-aggregated diagnostics (e.g., global averages), motivating efforts such as \citeA{prevost2025detection,rostron2025clearer} where the analysis of structural error is performed using regional or spatial clustering. This problem cannot be addressed by more efficient sampling methods because structural error is irreducible and will negatively impact parameter estimation results. Thus, the parameter estimation method should be able to account for the more frequent occurrence of structural errors, especially given the presence of emulator uncertainty, which we discuss below.

\subsection{Emulator complexity}
Including a larger number of parameters as emulator inputs can improve emulation accuracy for climate model PPEs \cite{yang2025simple}. However, this improvement comes at the cost of significantly increased dimensionality of the parameter space. In high-dimensional input spaces, different parameter combinations can compensate and yield nearly indistinguishable outputs, making it difficult to isolate the influence of individual parameters. As a result, some estimated parameters will appear poorly-constrained, although they are in fact better-constrained in the high-dimensional parameter space. This trade-off suggests that emulator designs that maximize emulator performance (i.e., more parameters as inputs) may not necessarily lead to more efficient parameter estimation results  \cite{chang2013fast}, which motivates the use of simpler emulators with fewer parameters as inputs. Examples of emulator designs utilizing the most sensitive parameters as inputs have been proposed in \citeA{williamson2015identifying,li2019reducing,rostron2025clearer}.

However, given many targets for parameter estimation as proposed in this work, it is likely that different groups of targets are sensitive to different but overlapping subsets of parameters. When these subsets share parameters, the resulting constrained regions in the parameter spaces across targets become coupled, and the effective dimensionality of the joint parameter space remains high. In such scenarios, although each emulator operates in a reduced input space (as we propose to reduce the number of input parameters for emulators), the calibration problem as a whole is still characterized by a high-dimensional parameter space, and the ill-posedness remains. Moreover, it is also difficult to explore the parameter space efficiently given finite samples when multiple targets jointly constrain the parameters to a highly localized region. This challenge serves as a larger obstacle as more calibration targets are included.

\subsection{The tolerance of structural error}
Introducing a tolerance for structural error is a practical compromise for parameter estimation, as model structural error can make it impossible to find any parameter combinations that match observations within some tolerance \cite{williamson2015identifying,elsaesser2025using,rostron2020impact}. The use of an emulator as an imperfect surrogate model adds further complication, as emulation uncertainty could exceed the magnitude of structural error in some situations. In other words, the values of certain emulated variables may be considered possible climate model outputs based on the emulated confidence interval, but those values are in fact unattainable by the climate model.

\begin{figure}[H]
    \noindent\includegraphics[width=\textwidth]{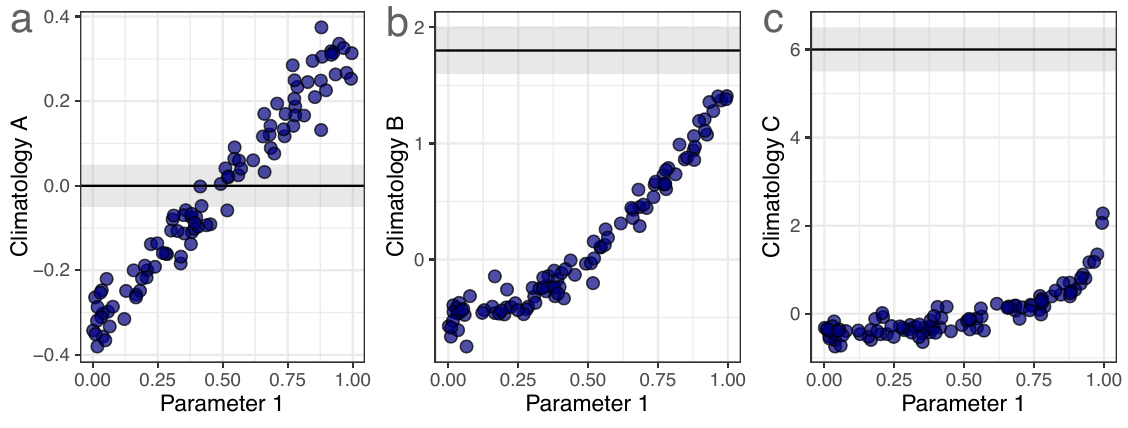}
    \caption{Different scenarios demonstrating the relationship between intrinsic structural error and observational constraints. Blue points denote model ensemble members from the PPE, and there are three variables of interest. Horizontal lines and gray shaded regions indicate the observational values and associated uncertainty ranges. The observations from Climatologies A, B, and C would individually constrain the parameter to approximately 0.5, 1.0, and 1.0, respectively. Structural error is present for both Climatologies B and C with the structural error being more severe for the latter (i.e., observation lies far outside the range spanned by all ensemble members). It should be considered intolerable for parameter estimation. If we are given observations of Climatologies A and B, a compromise that narrows the parameter to somewhere between 0.5 and 1.0 is acceptable. In contrast, given observations of Climatologies A and C, the latter should be excluded for parameter estimation, because otherwise the estimated parameters will be influenced by the observed Climatology C, but this influence will only degrade the performance for  Climatology A without significant improvement for Climatology C. }
    \label{f_structuralerror_example}
\end{figure}

In such cases, if we further increase the tolerance of the structural error, its impact is implicitly overestimated. The estimated parameters are likely biased towards the region in the parameter space where the structural error is stronger. A simple example is given in Fig. \ref{f_structuralerror_example}. For models that exhibit behaviors similar to that shown in Fig. \ref{f_structuralerror_example}, we propose avoiding the use of tolerances to compensate for structural error.

\subsubsection{Defining detectable structural error in the context of emulator-based parameter estimation}
Due to imperfect emulation, in the context of emulator-based model calibration, we can only partially identify structural error magnitude -- specifically, some quantitative mismatch estimate arising in situations where parameter values cannot produce model outputs consistent with observations, even after incorporating emulator predictive uncertainty and observational uncertainty. Assuming that the computed uncertainties are robust, this detectable structural error demonstrates the limitations in the model, not in the model parameter settings. In the following text, we use the term structural error to denote detectable structural error, as this is the primary form of structural error considered in our context. When we specifically refer to the actual, irreducible model structural error, we use the term intrinsic structural error.

\subsubsection{Treatment of detectable structural error}
We propose a somewhat counterintuitive strategy for handling structural error: when certain target variables exhibit structural error, the corresponding observations should be excluded from parameter estimation. Incorporating such variables can force the inferred parameters to compensate for model structural error that cannot be resolved through varying the parameter values (Fig. \ref{f_structuralerror_example}a and c).

Excluding variables with structural error does not necessarily lead to a loss of information. For example, if a regional diagnostic is affected by severe structural error, neighboring regions often exhibit weaker intrinsic structural error (and thus will not be excluded for calibration) and, due to spatial continuity, remain correlated with the affected variable. These neighboring diagnostics can still provide indirect, yet informative, constraints on the parameters.

\subsection{How many samples to draw}
Whether history matching or Bayesian methods are employed, parameter estimation relies on sampling from regions of the parameter space deemed consistent (to a certain level) with the observational data. In some scenarios, no parameter sets can be found or sampled. In addition to the presence of structural error discussed earlier, this may also occur when the parameters are constrained to an extremely small region by the observations, and a finite number of (emulated) samples could fail to fall within that region. In such scenarios, the absence of acceptable parameter samples does not necessarily indicate model structural error, but rather insufficient sampling density.

For example, if five parameters are restricted to 5$\%$ of their original ranges, the constrained region of the parameters is $0.05^5 \approx 3.1 \times 10^{-7}$ of its original size. In such a case, even if we generate one million candidate parameter samples using a rejection-based procedure (i.e., rejecting the samples if they do not fall within the constrained region in the parameter space), the expected number of samples that fall inside this small region remains below one. We are still likely to find no parameter samples due to the relatively sparse sampling density with the absence of structural error.

This example demonstrates that the impact of structural error and the practical challenge that is associated with the insufficient sampling density could co-exist for high-dimensional calibration problems. A calibration method should be able to unravel whether the absence of (acceptable) samples arises from model structural error or from the extremely well-constrained region in the parameter space with insufficient sample density.

\subsection{Automated or expert-informed calibration workflow}
Given the presence of structural error and its more frequent occurrence as the number and types of calibration targets increases, a growing number of subjective trade-offs will need to be considered during the parameter estimation process. These trade-offs cannot be resolved purely algorithmically and instead require expert judgment. For example, as demonstrated later in this work, increasing the value of a given parameter may improve the model skill for regional longwave cloud forcing (LWCF) while simultaneously degrading the performance for regional outgoing longwave radiation (OLR). In such scenarios, whether one should prioritize a particular variable, ignore another, or find a compromise depends on scientific objectives and domain expertise rather than the calibration method itself. A calibration workflow should be able to systematically reveal how parameters influence different variables and diagnostics and where structural error occurs, thereby providing critical information to support informed decision-making. Studies that implement sensitivity analyses, or that analyze aspects of structural errors, are increasing \cite{johnson2020robust,prevost2025detection}, and our approach integrates structural error detection and sensitivity analysis into the calibration workflow. The developed method enables both automatic (i.e., fully automated workflow) and manual selection of structural error-prone target variables during calibration.

\section{Method}
Our method design is based on the propositions listed in the previous section. The proposed workflow follows Fig. \ref{f_workflow} and is available on github (https://github.com/yiqioyang/proj2dhullsampler.git). Model variables open for calibration may include global averages, regional or seasonal climatologies, or other diagnostics such as extreme precipitation.  In contrast to some studies that tolerate structural error (\citeA{johnson2020robust,prevost2025detection,elsaesser2025using}), we sequentially identify and excludes target variables that bear structural error -- or in other words, we do not tolerate structural error. The method adopts an adaptive sampling strategy that identifies structural error while dynamically increasing the sampling density (i.e., the number of candidate samples) as the plausible region in the parameter space shrinks rapidly due to the constraints imposed by the observations.

Our proposed method is based on history matching \cite{williamson2013history}. The key technical innovation is that we decompose the overall calibration problem into a sequence of low-dimensional parameter estimation problems (that focus on subsets of the complete target variables),  and then integrate their information to identify parameter combinations that remain jointly consistent with all structural error-free observations.

We use Gaussian Processes (GPs) as emulators, training a separate GP for each variable of interest. Each GP is constructed using a selected pair of model parameters as inputs. The package Scikit-learn \cite{pedregosa2011scikit} is used to train the GPs in python.

In describing the method below, we assume that we are given the initial PPE or CPE which consists of a set of parameter samples and associated climate model outputs along with corresponding observations.

\subsection{Determining which parameters to use for emulator training}
The input parameters for each variable of interest are selected following a simple sensitivity test \cite{yang2025simple}. For each target variable, the method constructs a separate, independent 1D Gaussian process emulator for each of the parameters. These 1D emulators serve solely as a screening tool to assess parameter importance as part of emulator design, and they are not used in the parameter estimation process. The 1D emulator is deliberately regularized to avoid overfitting (i.e., fixed hyperparameters that will not lead to overfitted results during the training), as our objective is not optimal forward prediction but rather robust identification of dominant parameter influence.
Underfitting is therefore acceptable. The method then evaluates the 1D emulator performance using the root mean square error (RMSE) between emulated and simulated outputs. The two parameters associated with the lowest RMSE are identified as the most influential and are selected as inputs for training the GP emulator, with this final GP emulator serving as the surrogate model for parameter estimation. Here we restrain the number of input parameters to be less than three because the method is designed to work with sparse datasets. For example, with 100 samples (as done in this work), a two-dimensional grid achieves a resolution of 10 points per axis ($10 \times 10 = 100$). Under the same sample budget, a three-dimensional grid only reaches a resolution of roughly 5 points per axis ($5 \times 5 \times 5 = 125$). This coarsening illustrates how quickly the curse of dimensionality dilutes coverage of the parameter space as dimensionality increases, motivating our restriction to at most two input parameters.

\subsection{History matching and assessing structural error for target variables}
After the emulators are trained for each variable of interest, we use 
Latin Hybercube Sampling (LHS) to sample $n$ parameter vectors $\mathbf{x}_i \in \mathbb{R}^m$, $i = 1,\dots,n$ uniformly based on the pre-specified parameter ranges (which are defined based on the parameter ranges of the original PPE or CPE). Here $m$ denotes the number of model parameters, and we write
\[
\mathbf{x}_i = (x_{i,1}, x_{i,2}, \dots, x_{i,m}).
\]
Throughout this work we set $n=10^6$.  

As mentioned earlier, each variable of interest $y^k$ ($k=1,\dots,K$) is associated with a $GP^k(\cdot)$ emulator constructed using its two most sensitive parameters as inputs. We denote $S_k \subset \{1,\dots,m\}$ as the indices of these sensitive parameters for variable $y^k$, and define
\[
\mathbf{x}_i^k = (x_{i,j})_{j \in S_k},
\]
such that $\mathbf{x}_i^k$ corresponds to the two-parameter inputs to the emulator for $y^k$ for sample $\mathbf{x}_i$. The GP prediction for sample $\mathbf{x}_i$ can be written as $GP^k(\mathbf{x}_i^k)$, with predictive mean and standard deviation denoted as $\tilde{\mu}_i^k$ and $\tilde{\sigma}_i^k$, respectively. The sampled parameters together with their corresponding emulated variables constitute the emulated ensemble.

Let $y^k_{\mathrm{obs}}$ be the observed value of variable $y^k$, with observational uncertainty characterized by standard deviation $\sigma^k_{\mathrm{obs}}$. We define a binary plausibility indicator for each sampled parameter vector $\mathbf{x}_i^k$ as

\begin{equation}
I^k(\mathbf{x}_i^k) = 
\begin{cases}
1, & \tilde{\mu}_i^k - \tau \sqrt{(\tilde{\sigma}_i^k)^2 + (\sigma^k_{\mathrm{obs}})^2}
\le y^k_{\mathrm{obs}} \le
\tilde{\mu}_i^k + \tau \sqrt{(\tilde{\sigma}_i^k)^2 + (\sigma^k_{\mathrm{obs}})^2}, \\
0, & \text{otherwise},
\end{cases}
\label{eq_plausible}
\end{equation}
where $\tau$ is a scaling factor controlling the acceptance width.

When observational uncertainty is provided as an admissible interval $[y^k_{\mathrm{obs,min}},\, y^k_{\mathrm{obs,max}}]$, the criterion is similarly written as

\begin{equation}
I^k(\mathbf{x}_i^k) =
\begin{cases}
1, & \max\!\left(\tilde{\mu}_i^k - \tau \tilde{\sigma}_i^k,\; y^k_{\mathrm{obs,min}}\right)
<
\min\!\left(\tilde{\mu}_i^k + \tau \tilde{\sigma}_i^k,\; y^k_{\mathrm{obs,max}}\right), \\
0, & \text{otherwise}.
\end{cases}
\label{eq_plausible_b}
\end{equation}
Eq. \ref{eq_plausible} assigns plausibility when the observational mean lies within a $\tau$-scaled interval which is defined by the observational and emulator uncertainty. Here the emulator and observational uncertainty is considered independent of each other. The term $\sqrt{(\tilde{\sigma}_i^k)^2 + (\sigma^k_{\mathrm{obs}})^2}$ in Eq. \ref{eq_plausible} corresponds to the standard deviation of the sum of two independent Gaussian random variables, namely the emulated variable and the zero-mean observational noise. Eq.\ref{eq_plausible_b} requires a nonempty intersection between the emulator uncertainty interval and the observational admissible range.

Since we do not explicitly tolerate structural error in this framework, we set $\tau=2.0$ throughout this work unless otherwise stated, corresponding approximately to the 95$\%$ confidence interval for Gaussian random variables. Sensitivity to this setting is tested later in this work. Increasing $\tau$ to a value greater than 2.0 is acceptable in practice, but the justifications should be related to the quality and size of the dataset (e.g., given 40 samples, we need to increase $\tau$ because the predicted emulator uncertainty tends to be less robust) instead of trying to tolerate more structural error. The method also has a validation step that checks how many training samples (i.e., the PPE or CPE data) are outside of the prediction uncertainty range (defined by the predicted mean, standard deviation, and specified $\tau$). Users can set a threshold on it such that the variables with poor-performance emulators will be excluded for parameter estimation. Throughout this work, this threshold is set to be 10$\%$, and is never triggered.

Ideally, parameter vector $\mathbf{x}_i^k$ satisfying
\[
\sum_{k=1}^K I^k(\mathbf{x}_i^k) = K
\]
would be retained as plausible parameter estimates. In practice, this is almost impossible when a large number (e.g., more than 20) of calibration targets are used, due to the presence of structural errors across variables and our design choice that we do not tolerate the structural errors.

We therefore introduce a user-selectable, pre-filtering process that excludes variables of interest that satisfy:
\begin{equation}
\sum_{i=1}^n I^k(\mathbf{x}_i^k) < n\phi,
\label{eq_phi}
\end{equation}
where $\phi \in [0,1]$ is a prescribed threshold. Variables $y^k$ with $\sum_{i=1}^n I^k(\mathbf{x}_i^k)$ being small or zero constrain their sensitive parameters to a very small region or yield no surviving parameter samples. When many targets are considered, such variables, especially when they are non-global, likely bear strong structural error and are excluded. In practice, we suggest $\phi$ to be in the range of 0.00-0.20. Namely, if one (non-global) observation alone could constrain the parameters to less than $0\%-20\%$ of their originally specified sample size (i.e., the ratio of surviving samples satisfying $I^k(\mathbf{x}_i^k) = 1$ and the original sample size $n$), then it is likely that this observation and the corresponding variable bear strong structural error. The exclusion of such a non-global variable could potentially lead to a waste of information. The justifications are: (1) Variables can be spatially correlated, even with this variable excluded, it is correlated with the neighboring variables, which could play similar roles in helping constrain the parameters; (2) The local variables may bear strong structural error. Constraining the parameters strongly with just one such variable may intensify the biases elsewhere or even globally. We also note that this is a non-mandatory step for the method, given the possibility of one important observation constraining to a small portion of the parameter space, and by setting $\phi$ to zero, this step of excluding the variables is skipped. We also note here that the exclusion of the target variables can be viewed as a pre-screening process. Even if we set the value of $\phi$ to 0.0, i.e., not excluding any variable at this stage, the variables that bear strong structural error will be identified and excluded for calibration in later stages of the method.

\begin{figure}[H]
    \noindent\includegraphics[width=\textwidth]{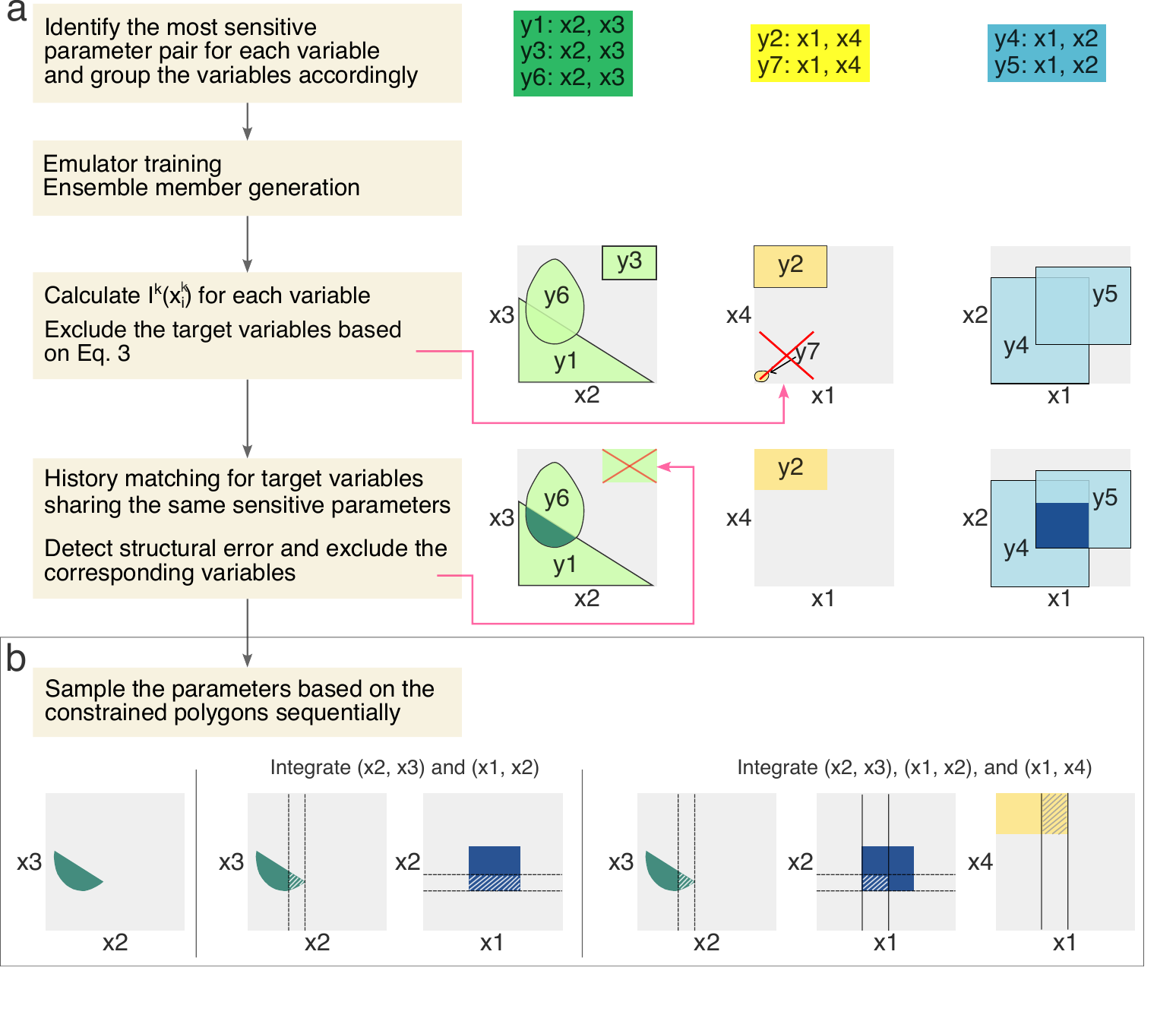}
    \caption{Proposed method workflow. The xs and ys correspond to model parameters and target variables, respectively. a: different variables are grouped based on their most sensitive parameters, and history matching and exclusion of variables are implemented to constrain the parameters; b: how the constrained parameters are integrated through sequential sampling. The shaded areas correspond to the constrained parameters. }
    \label{f_workflow}
\end{figure}

The remaining variables are grouped according to their sensitive parameter pairs (Fig. \ref{f_workflow}a). This procedure is similar to \cite{prevost2025detection}, but we do not further aggregate them to devise new target variables and observations. For each group, we examine whether there exist parameter samples $\mathbf{x}_i^k$ satisfying $I^k(\mathbf{x}_i^k)=1$ for all variables in the group. If no such sample exists, structural error exists for variables within this group (i.e., the variables that share the same parameter pair as their sensitive parameters). We need to identify and exclude the variables that lead to the structural error.

To do so, we iteratively consider all pairwise combinations of variables within each group that has structural errors and share the same sensitive parameters, and compute the number of surviving samples satisfying plausibility conditions for both variables (i.e., $I^k(\mathbf{x}_i^k)=1$ for the two variables). Variable pairs that jointly and consistently lead to zero or near-zero occurrence of plausible samples are identified as those associated with the strongest structural errors and are subsequently excluded from being used for calibration (Fig. \ref{f_workflow}a). Here the threshold for exclusion can be specified by the users. We set its default value to be identical to that of $\phi$.

\subsubsection{Two ways to exclude structural error-laden variables}
By default, the method automatically removes such variables one at a time within each group of variables. The variables are chosen in the following way: we count the occurrence of variable pairs that lead to zero or near-zero plausible samples. The variable with the greatest count is assumed to have the strongest structural error (Table \ref{t_strt_err}). After this variable is excluded, we iterate this procedure until the remaining variables do not lead to zero or near-zero amount of plausible samples when considered jointly.

Alternatively, the method allows users to specify the variables to be excluded based on their expert judgment. The decision can be further informed by the summary of how different variables contribute to structural error (e.g., Table \ref{t_strt_err}). In such cases, users should be certain that the remaining variables do not lead to zero or near-zero occurrence of plausible samples when considered jointly.

When pairwise combinations are insufficient to identify the source of inconsistency, we can extend the analysis to higher-order combinations of three or more variables. This can be done by the method, but we have not found this necessary in practice.

After applying the above procedure across all 
groups of variables that exhibit structural errors, certain variables are excluded due to the presence of structural error. For each group of remaining target variables (that share the same sensitive parameters), no structural error exists in their respective two-dimensional subspaces.

\subsubsection{Defining the remaining plausible region as a polygon}
For each sensitive parameter pair $(p_a,p_b)$, we collect all parameter samples $\mathbf{x}_i$ that satisfy the plausibility criteria for every variable associated with that pair (i.e., $I^k(\mathbf{x}_i^k) = 1$ for all $y^k$ in the group and are not excluded in previous steps), and extract their projected coordinates $(x_{i,a},x_{i,b})$. These surviving projected points approximate the plausible region in the corresponding $(p_a,p_b)$ subspace. Using these projected points, the plausible region is constructed using an alpha-shape method (based on the Alpha Shape Toolbox; \citeA{bellock2019alphashape}), denoted as $\mathcal{H}_{ab}$ for parameter pair $(p_a,p_b)$. For each two-dimensional parameter subspace, its corresponding plausible region can be viewed as a polygon filled with the surviving samples.

However, the parameter pairs are not independent of each other. We could have one parameter pair, a and b, and another, b and c. We do not know whether structural error exists for the joint parameter space of \{a, b, c\}. When several more pairs of parameters are inter-linked in this way, the dimensionality of the parameter space increases, and structural error may still exist.

\subsection{Integrating pairwise constraints to the full parameter space}
To explore consistency across linked parameter pairs, we propagate the low-dimensional constraints into the full parameter space using a sequential rejection-sampling procedure. We generate another ensemble (e.g., $10^5$) of candidate parameter samples $\mathbf{x}_i$ using Latin hypercube sampling (LHS) within the pre-specified parameter ranges. For each parameter sample, we extract its two-dimensional projections $(x_{i,a},x_{i,b})$ corresponding to each sensitive parameter pair $(p_a,p_b)$. We then test sequentially whether each projected point lies within the plausible region $\mathcal{H}_{ab}$. If a candidate sample is outside $\mathcal{H}_{ab}$ in the projected subspace, it is discarded. This filtering process is applied sequentially across all plausible regions of different parameter pairs, progressively restricting the set of plausible parameter samples (Fig. \ref{f_workflow}b).

As more plausible regions are used to constrain the samples, the plausible region typically shrinks rapidly. In some cases, the surviving sample set becomes empty after enforcing a particular constraint, for example, $\mathcal{H}_{ae}$ for parameter pair $(p_a,p_e)$. As mentioned earlier, this can be explained by two scenarios: (i) insufficient sampling density in a highly localized and constrained region in the full parameter space, or (ii) structural error tends to occur more often when more (regional) targets are considered.

To distinguish between these two possibilities, we restart the sequential rejection-sampling procedure using a substantially increased number of candidate samples, typically by two orders of magnitude. If surviving parameter samples emerge under this denser sampling, the earlier presence of the empty set is attributed to insufficient sample size. The method then proceeds with the increased sampling size. Otherwise, if no samples survive even with denser sampling, we interpret this as evidence of structural error arising from the joint interaction of multiple linked parameter pairs. We need to further examine whether the conflict is driven by only a subset of variables within the corresponding parameter-pair group, rather than requiring the entire group to be discarded.

To be consistent with the previous example, let $\mathcal{V}_{ae}$ denote the set of target variables whose sensitive parameters are $(p_a,p_e)$. We iteratively rebuild the plausible region for $(p_a,p_e)$ by excluding one or more targets from $\mathcal{V}_{ae}$, and test whether any parameter samples could satisfy this updated constraint together with all previously considered constraints (i.e., those defined by other parameter pairs). If such parameter samples exist when certain targets are excluded, we identify them as the contributors to the structural error. The plausible region for $(p_a,p_e)$ is then updated based on the remaining variables within the group $\mathcal{V}_{ae}$, and the sequential filtering procedure resumes with the remaining parameter pairs and with the increased sampling size. If no parameter samples can be found for any combination of variables, we conclude that the structural error cannot be addressed by excluding a subset of the target variables within the group. In this case, the entire group (i.e., $\mathcal{V}_{ae}$) of targets associated with $(p_a,p_e)$ is removed from further consideration.

Our method effectively propagates low-dimensional constraints into higher-dimensional parameter space while avoiding direct emulation in high dimensions. The resulting, surviving parameter samples which are consistent with all retained calibration targets provide the final estimate from the method.

Because parameter-pair constraints are applied sequentially and the targets are excluded sequentially, the ordering of the sequence can affect the resulting parameter samples. Such sequences can be rearranged in the method as needed to assess robustness or detect other plausible regions in the parameter space.

\subsection{Practical concerns and limitations}

The three-stage automated method yields excluded variables and their sensitive parameter pairs, and this information is recorded and output automatically for users. An example of the model outputs is presented in Appendix A. The major limitation of the proposed method is that all remaining variables used for parameter estimation are considered equally important. This choice is intentional, as assigning and estimating variable-specific weights as done in conventional Bayesian parameter estimation will become impractical given many targets, especially in the presence of structural errors. Additionally, the method design is also ``greedy'' in maximizing the extent to which parameters can be constrained. It faces the risk of over- and incorrect-constraining, which will be demonstrated in later experiments. 

Although our method is set up to use non-aggregated diagnostics for parameter estimation, it is still practical and likely beneficial to have additional globally-averaged diagnostics as targets for parameter estimation. In such a situation, we simply do not consider excluding such variables, thereby ensuring for example that global constraints such as the energy balance are considered in estimating the parameters.

\section{Application of the method to the Community Atmosphere Model (CAM)}
We use a version of CAM6 in which the warm rain microphysics parameterization (i.e., autoconversion and accretion) is replaced with a machine learning-based model (hereafter, CAM6-ML). This Machine Learning-based parameterization is similar to \cite{gettelman2021machine} but trained with a simpler neural network structure. The training data for the parameterization is based on a version of CAM6 in which warm rain microphysics is represented with a bin-resolved scheme that more explicitly represents the processes. There is no systematic prior knowledge on how this Machine Learning-based parameterization affects the simulated climate. We use CAM6-ML to generate a PPE following the approach used for the existing CAM6 PPE \cite{eidhammer2024extensible}. We use Latin Hybercube Sampling (LHS) to generate 100 parameter sets with 34 parameters varied. The parameters span the microphysics, deep convection, turbulence, and aerosol parameterizations. Their ranges are determined with expert elicitation and are given in Table \ref{t_para_descrip}.

For each simulation, the near present day cyclic boundary conditions representative of the year 2000 are used. The greenhouse gases and atmospheric oxidants take the average values of the 1995-2005 period. The average monthly sea surface temperatures (SSTs) during 1995-2010 are used. The emission of aerosols and precursors is set to the monthly 1995-2005 climatology in these simulations. Each simulation runs for a period of 3 years.

\begin{figure}[H]
    \noindent\includegraphics[width=\textwidth]{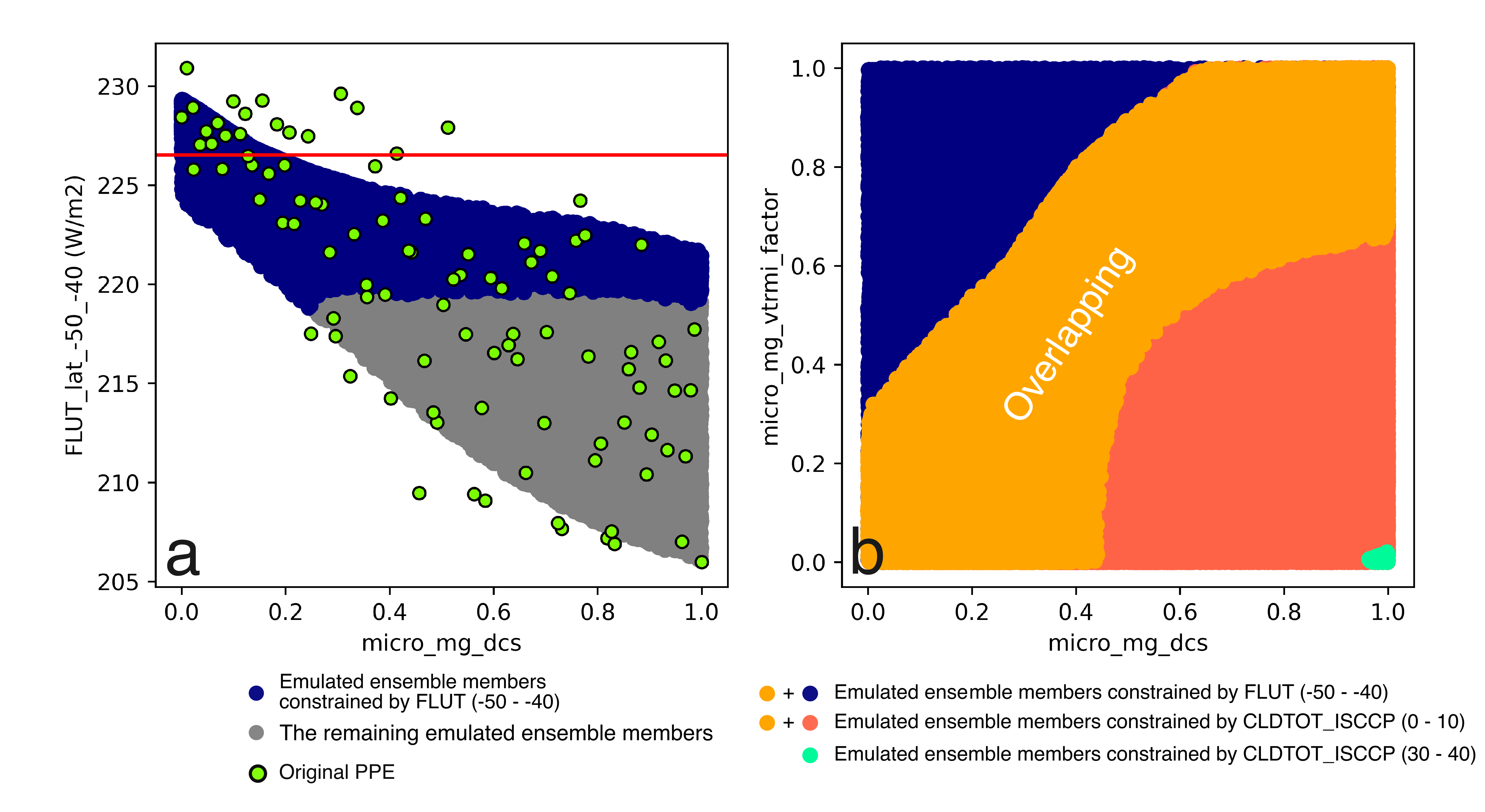}
    \caption{a: Example of how the history matching is implemented for one target diagnostic. Green points: FLUT from $-50^\circ$S to $-40^\circ$S plotted against its most sensitive parameter \textit{micro$\_$mg$\_$dcs} from the original PPE. Blue and gray point clouds correspond to the mean of the emulated ensemble members with the blue points having $I^k(\mathbf{x}_i^k) = 1$ for this variable; b: the constrained ensemble members from three different variables sharing the same sensitive parameter pair \textit{micro$\_$mg$\_$dcs} (x-axis) and \textit{micro$\_$mg$\_$vtrmi$\_$factor} (y-axis). Note that the orange points correspond to the ensemble members that satisfy $I^k(\mathbf{x}_i^k) = 1$ for both FLUT from $-50^\circ$S to $-40^\circ$S and CLDTOT$\_$ISCCP from $0^\circ$ to 10$^\circ$N. Note that the ensemble members constrained using CLDTOT$\_$ISCCP from $30^\circ$N to 40$^\circ$N are at the lower right corner in b (green). }
    \label{f_constrain}
\end{figure}

Our target climatologies and their corresponding observational products are provided in Table \ref{t_obs_descrip} (\citeA{huffman2007trmm,elsaesser2017multisensor,loeb2018clouds,adler2018global,rossow2022international}), and most of the individual products were downloaded originally from obs4MIPs \cite{waliser2020observations}. Observational uncertainty is assumed to be zero which is sufficient for establishing proof of concept in this work. The targets focus on the top-of-atmosphere energy budget (including cloud effects) and the hydrologic cycle (precipitation and precipitable water vapor). They are frequently used in traditional climate model tuning. We do not use cloud liquid water path (TGCLDLWP) and total cloud fraction (CLDTOT$\_$ISCCP) to constrain the observations due to their well-known and strong structural error for CAM (e.g., \citeA{kay2012exposing,medeiros2023assessing}). Considering them in the parameter estimation would significantly degrade the model skill. We set our target variables to be the climatologies averaged over discrete latitude bands (e.g., 5$^\circ$ or 10$^\circ$ bands). For example, LWCF\_lat\_45\_50 denotes LWCF averaged over time and the latitude band from 45$^\circ$N to 50$^\circ$N. This notation is used throughout this work. We note here that the spatial averaging indeed compresses the spatial information which we do not support, but the resulting target variables still contain significantly more information than simple global means or metrics. We therefore view the use of the target variables defined above as a representative yet tractable test case for demonstrating the proposed framework. Fig. \ref{f_constrain}a shows the relationship between FLUT from 50$^\circ$S to 40$^\circ$S and its most sensitive parameters, and how the emulated ensemble members are constrained by the corresponding observation. Fig. \ref{f_constrain}b shows how three different variables constrain the parameter space. They share the same sensitive parameters. It is noted that the observed CLDTOT$\_$ISCCP from 30$^\circ$ to 40$^\circ$ constrains the parameters to a very small region in the lower-right corner of Fig. \ref{f_constrain}b. This highlights the strong structural error, and helps justify the exclusion of TGCLDLWP and CLDTOT$\_$ISCCP for parameter estimation.

In addition to the 100-member PPE, we also run the CAM6-ML using the CAM6 default parameters, referred as the CAM6-ML default run. It is not used for training the emulators, but used as a reference for performance evaluation together with satellite observations and the CAM6 default run which does not use the Machine-Learning based parameterization (i.e., the CAM6 default run with its default parameters).  It is out of scope to compare CAM6 and CAM6-ML in this paper, but we refer the readers to Table 4 for a brief summary of the differences across a number of diagnostics simply for reference.

Global biases and zonal climatologies for the CAM6-ML PPE are presented in Figs. \ref{f_global_bias} and \ref{f_zonal_bands}. It is noted that none of the original PPE members could match the observed LWCF and FLUT global averages at the same time (specifically Fig. \ref{f_global_bias}, where no ensemble member has near-zero bias for both). This represents a structural error that cannot be resolved by varying the values of the parameters.

\section{Results}
Following the developed method and the CPE workflow \cite{elsaesser2025using}, we evaluate the CAM results and test impacts of various thresholds. These tests are reported as ``iterations'' and are described below.

\subsection{Iteration One design and results}

Since the method is designed to work in the presence of structural error, we focus on testing the method with real observations. However, first, to demonstrate the proof of concept, we choose one simulation to serve as the ``observation'' and the remaining simulations as input to the method. The experiment is repeated 10 times with different simulations chosen as the observational reference. The results are presented in Appendix B. None of the true parameter values are outside the sampled parameter ranges, and the parameters are constrained to 2$\%$ to 62$\%$ of the original size. These results suggest that the method is able to effectively and correctly constrain the parameters.

We apply the method as introduced to the original 100-member PPE, defined here as Iteration 1. Our expectation for this iteration is to correctly constrain the parameters such that the large spread in the original PPE can be reduced (Fig. \ref{f_zonal_bands}). We do not attempt to find optimal ensemble members in this iteration. This task is left to subsequent iterations introduced later.

We set the threshold of $\tau$ in Eq. \ref{eq_plausible} to 2.0. Our target variables are the climatologies listed in Table \ref{t_obs_descrip} with the exclusion of TGCLDLWP and CLDTOT$\_$ISCCP, averaged over 10$^\circ$ latitude bands within the range of  -70$^\circ$S to 70$^\circ$N. We do not include globally averaged quantities, as the latitude bands between -70$^\circ$S and 70$^\circ$N already cover the majority of the surface of the earth and capture the dominant spatial variability relevant for parameter estimation. Latitude bands south of -70$^\circ$S and north of 70$^\circ$N are excluded because variability in these polar regions is comparatively small and contributes limited additional constraint to the parameters in the present framework where the ocean and ice are inactive.

Using a threshold of 0.1 for $\phi$ defined in Eq. \ref{eq_phi}, we exclude two target variables, namely, TMQ from -70$^\circ$S to -60$^\circ$S and PRECT from -70$^\circ$S to -60$^\circ$S in Iteration 1.

We draw 90 parameter samples from the proposed method, and run the CAM6-ML accordingly. We are left with 87 simulations after three failed. They are referred to as the Iteration 1 ensemble. The Iteration 1 envelopes defined by the min-max spread of its zonal climatologies are presented in Fig. \ref{f_zonal_bands}. Compared to the original 100-member PPE, the spread is significantly reduced, especially near the equator for SWCF, LWCF, FLUT, and FSNTOA.

The global biases are shown in Fig. \ref{f_global_bias}.  Despite the structural error between FLUT and LWCF noted earlier, the Iteration 1 global biases are in general closer to zero than those of the original PPE.

Using the product of the normalized parameter ranges as an approximation, the constrained region in the parameter space after Iteration 1 shrinks to 0.29$\%$ of its original size. The normalized parameter ranges from the Iteration 1 ensemble together with biplots of selected parameter pairs are presented in Fig. \ref{f_shrink_cor}.
The strongly constrained parameters include \textit{micro$\_$mg$\_$vtrmi$\_$factor}, \textit{micro$\_$mg$\_$berg$\_$eff$\_$fact}, \textit{micro$\_$mg$\_$dcs}, \textit{clubb$\_$c1}, \textit{zmconv$\_$tiedke$\_$add}, and \textit{zmconv$\_$capelmt}. Among these parameters, some of their corresponding CAM6 default values are outside the constrained ranges (e.g., \textit{micro$\_$mg$\_$vtrmi$\_$factor}). This is due to the version difference between CAM6 and the CAM6-ML that is used in generating the PPE.

\subsection{Iteration Two}
\subsubsection{Design}
We apply the proposed method to the Iteration 1 ensemble with different designs and goals. We refer to them as Iterations 2a, 2b, and 2c. This iterative application of the parameter estimation method follows
\citeA{williamson2013history,elsaesser2025using}. We compute zonally-averaged climatologies with a latitudinal resolution of 5$^\circ$ from -75$^\circ$S to 75$^\circ$N and use them as targets for the three iterations. We increase the resolution here (i.e., from 10$^\circ$ in Iteration 1 to 5$^\circ$ here) with the expectation that this finer spatial averaging provides additional information to better constrain the parameters.

The three iterations are designed in the following way: in Iteration 2a, we set $\tau = 2.0$ and $\phi = 0.05$; Iteration 2b is identical to Iteration 2a except that we completely exclude LWCF for parameter estimation; in Iteration 2c, we set $\tau = 2.5$ and $\phi = 0.01$.

We design Iteration 2a with the expectations that 1) the sampled parameters would lead to better constrained zonal climatologies compared to those of Iteration 1; and 2) we might be able to find one or two sets of parameters that lead to overall comparable or better model skills compared to the CAM6-ML default run and potentially the CAM6 default run. We design Iteration 2b as an example of removing known structural error between FLUT and LWCF, as noted earlier (Fig. \ref{f_global_bias}e). The Iteration 2b design is expected to improve the model skill for FLUT at the cost of LWCF.

We design Iteration 2c to illustrate how increasing $\tau$ could impact the parameter estimation. Here the increased $\tau$ could be interpreted as either (1) a wider uncertainty estimate or (2) the introduction of tolerance on structural error, which our method aims to avoid, and which we deliberately construct here to illustrate its consequences. In the former interpretation, the increase of $\tau$ from 2.0 to 2.5 is not an overestimation of the uncertainty, because 99.7$\%$ of a Gaussian distribution (which is the output of the emulator) falls within three standard deviations of the mean (the three-sigma rule). In the latter interpretation, we increase the tolerance for structural error uniformly for all target variables, rather than varying $\tau$ per variable (in practice, this should be variable dependent). This uniform increase is more consistent with existing approaches. Previous works tend to inflate tolerance for structural error after aggregating diagnostics into cluster means or regional/global averages. This can look like a selective, variable-specific adjustment, but within each aggregated target, the underlying, unaggregated and spatially-varying diagnostics still receive the same undifferentiated tolerance once grouped together. Because our method instead retains targets at the individual latitude-band level, we keep $\tau$ constant for all variables uniformly.

\begin{figure}[H]
    \noindent\includegraphics[width=0.8\textwidth]{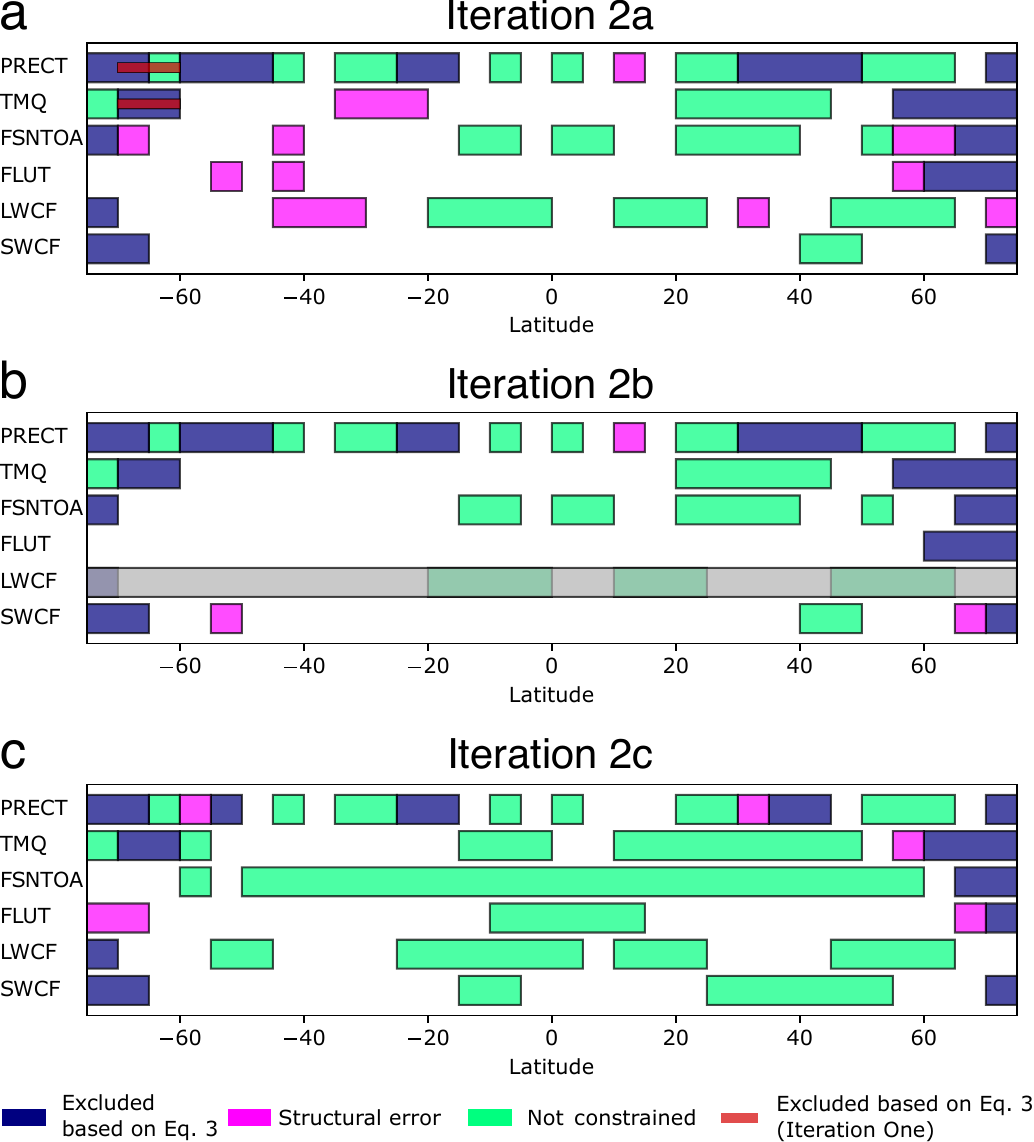}
    \caption{Target variables that are not used for constraining the parameters in Iterations 2a, 2b, and 2c are categorized and color coded. The climatologies that are excluded in Iteration 1 are also marked in a. Note all LWCF are not considered in Iteration 2b (b) which is by design.}
    \label{f_latband}
\end{figure}

We note that when the proposed method is used, there are also target variables whose observations do not constrain the parameters (i.e., $y^k$ with $\sum_{i=1}^n I^k(\mathbf{x}_i^k) = n$). These variables, together with the ones that are excluded during the implementation of the method, are color-coded according to their roles in Fig. \ref{f_latband} for Iterations 2a-2c. Due to the complete exclusion of LWCF, a substantially larger number of FLUT, FSNTOA, and TMQ variables are used to constrain the parameters in Iteration 2b relative to Iteration 2a. In Iteration 2c, $\tau$ is increased and as a result, substantially fewer target variables are utilized to constrain the parameters, newly expanding to include some variables that were previously providing stronger constraint (i.e., green bars in Fig. \ref{f_latband}c); this happens because of the imposed uniform increase in $\tau$.

Following the different designs, we draw samples from Iterations 2a, 2b, and 2c using the proposed method. We draw $90$ parameter samples from each iteration, and run them in CAM6-ML. We evaluate performance for each model configuration from four perspectives: zonal climatology comparisons; the global RMSEs and biases with respect to the observations; the performance of the best ensemble members; and the size of the constrained region in the parameter space.

\subsubsection{Zonal climatologies}
The Iteration 2a zonal climatology ensembles are, in most cases, better constrained to observations than those of Iteration 1. For some zonal bands the observed PRECT is outside the envelope. Additionally, the observed FLUT is slightly higher than the upper bound of the envelope for the most poleward latitudes.

\begin{figure}[H]
   \noindent\includegraphics[width=\textwidth]{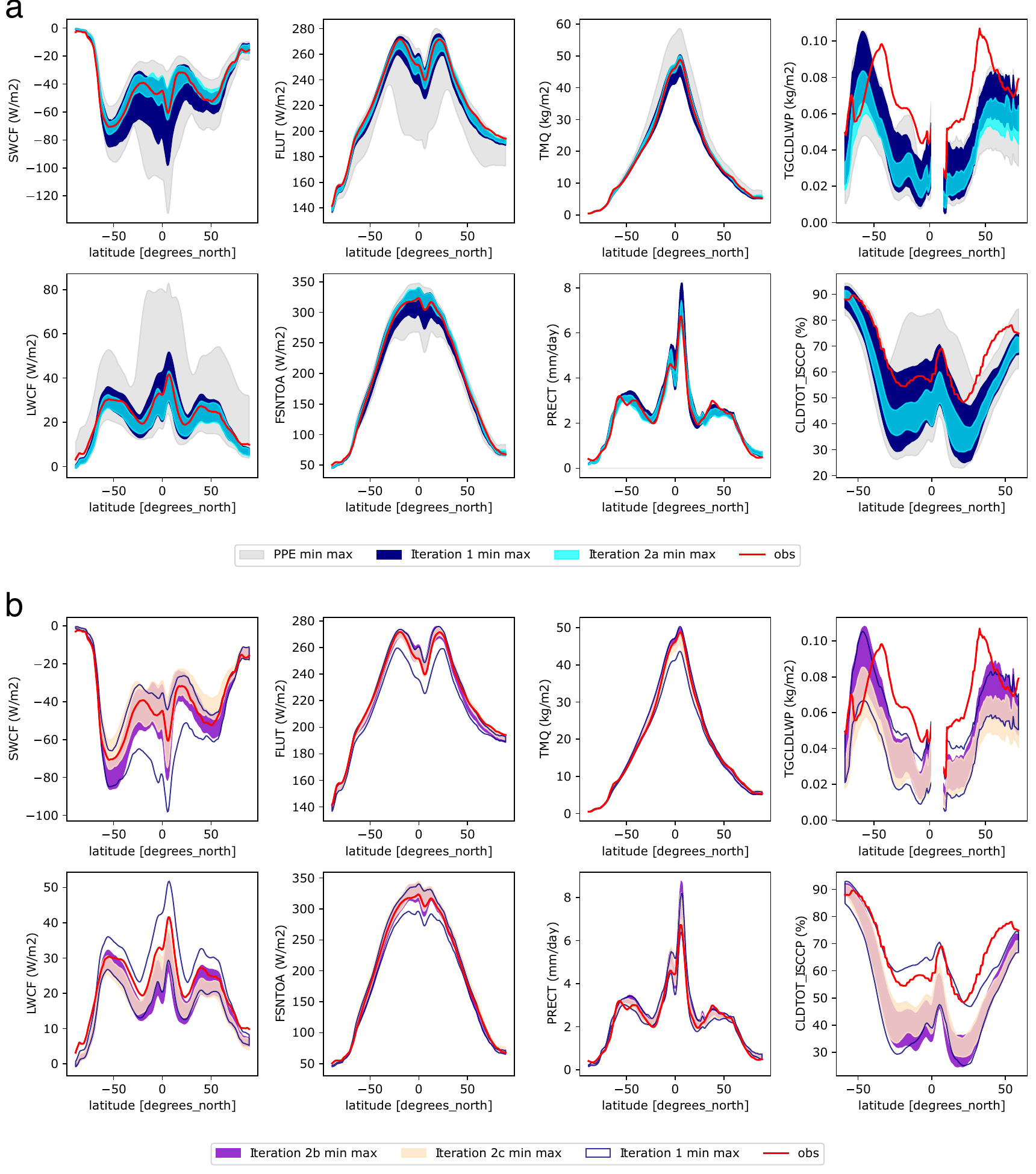}
    \caption{a: Minimum-maximum zonal climatology envelopes of the original PPE, and Iterations 1 and 2a ensemble members; b: minimum-maximum zonal climatology envelopes of the Iterations 1, 2b, and 2c ensembles. The observations are marked as red lines. The envelopes of Iteration 1 is marked as solid lines in b for reference. }
    \label{f_zonal_bands}
\end{figure}

The Iteration 2b ensemble envelopes are generally more constrained to observations with the exception of LWCF. The observed LWCF exceeds the Iteration 2b envelope upper-bound in many latitude bands (e.g., -40$^\circ$S to 0$^\circ$). It is also noted that FLUT is better constrained compared to the Iteration 2a envelope (Fig. \ref{f_zonal_bands}).

The Iteration 2c ensemble envelopes are similar to that of the Iteration 2b ensemble but with even more locally degraded performance for LWCF. Interestingly, with increased $\tau$ and decreased $\phi$ (which can be viewed as introducing more tolerance on the structural error), the LWCF envelope of Iteration 2c near 50$^\circ$N is farther away from the observation compared to that of Iteration 2b (Fig. \ref{f_zonal_bands}). This will be analyzed in more detail later in the text. Except for this observation, the envelope comparison between Iteration 2a, 2b, and 2c presented in Fig. \ref{f_zonal_bands} meet our design expectations and hypotheses.

\subsubsection{Global bias}
\begin{figure}[H]
    \noindent\includegraphics[width=\textwidth]
    {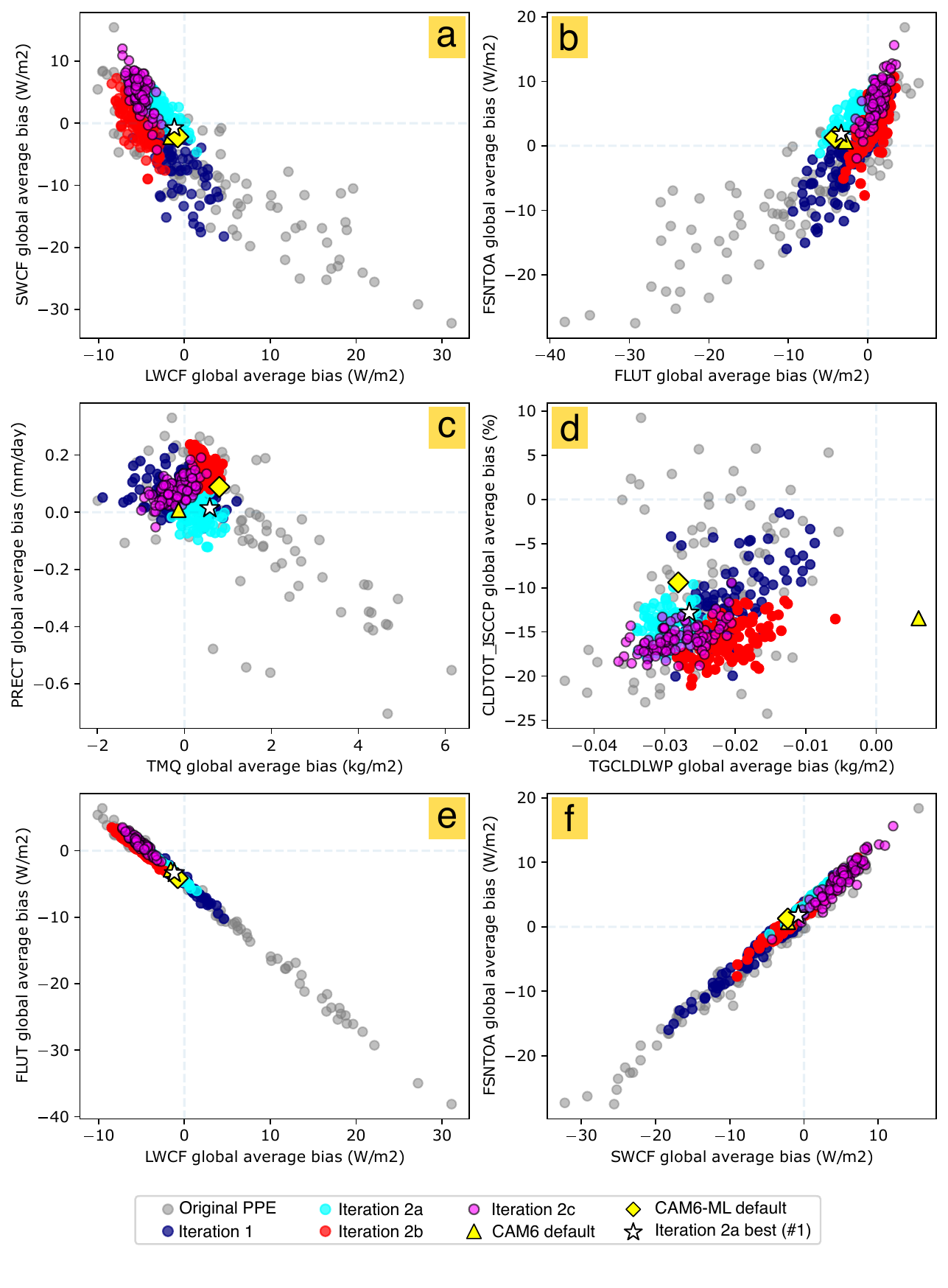}
    \caption{Global biases of the target climatologies from the original PPE and Iteration 1 and Iterations 2a, 2b, and 2c ensembles. FLUT and LWCF biases are plotted in e and FSNTOA and SWCF biases are plotted in f.  They are plotted to highlight the structural error of LWCF and FLUT. }
    \label{f_global_bias}
\end{figure}

The Iteration 2a-2c ensembles' global-average biases are plotted in pairs in Fig. \ref{f_global_bias}a-d. Ensemble members from Iteration 2a (cyan) are characterized by lower-magnitude LWCF, SWCF, and PRECT biases compared to those from Iteration 1. Ensemble members from Iteration 2b (red) have their FLUT and FSNTOA biases closer to zero at the cost of worse LWCF and PRECT. This is consistent with the intended impact of the design. The global biases for the Iteration 2c ensemble members (purple) lie between the Iteration 2a and 2b point clouds except for TMQ, where biases are on average smaller than those of the other two ensembles. The Iteration 2c ensemble resembles the Iteration 2b ensemble more in FLUT and LWCF, but resembles Iteration 2a more in SWCF and FSNTOA.

Fig. \ref{f_global_bias} also reveals an implicit structural error that is difficult to visualize in 2D parameter-output plots. The Iteration 2b ensemble in Fig. \ref{f_global_bias}c is characterized by greater bias in PRECT. Together with its reduced FSNTOA and FLUT biases (as well as comparison with the Iteration 2a ensemble), this shows that improvement in the simulation of FSNTOA and FLUT comes at the cost of degraded performance in PRECT simulation.

\subsubsection{Best ensemble members}
We calculate the RMSE of the ensemble members from Iterations 1 and 2a-2c. From each iteration, we pick two ensemble members characterized by relatively low RMSE and bias in all climatologies (Table \ref{t_rmse}). The two best ensemble members (\#1 and \#59) from Iteration 2a exhibit lower RMSEs than the CAM6 default run for all climatologies. Ensemble member \#1 from Iteration 2a also outperforms the CAM6-ML default run in RMSE for all climatologies but CLDTOT$\_$ISCCP. The best ensemble members from Iterations 2b and 2c exhibit SWCF and FLUT RMSEs lower than those of the CAM6 default run; however, the LWCF, FSNTOA, TMQ and PRECT RMSEs have increased overall.

\begin{figure}[H]
   \noindent\includegraphics[width=\textwidth]{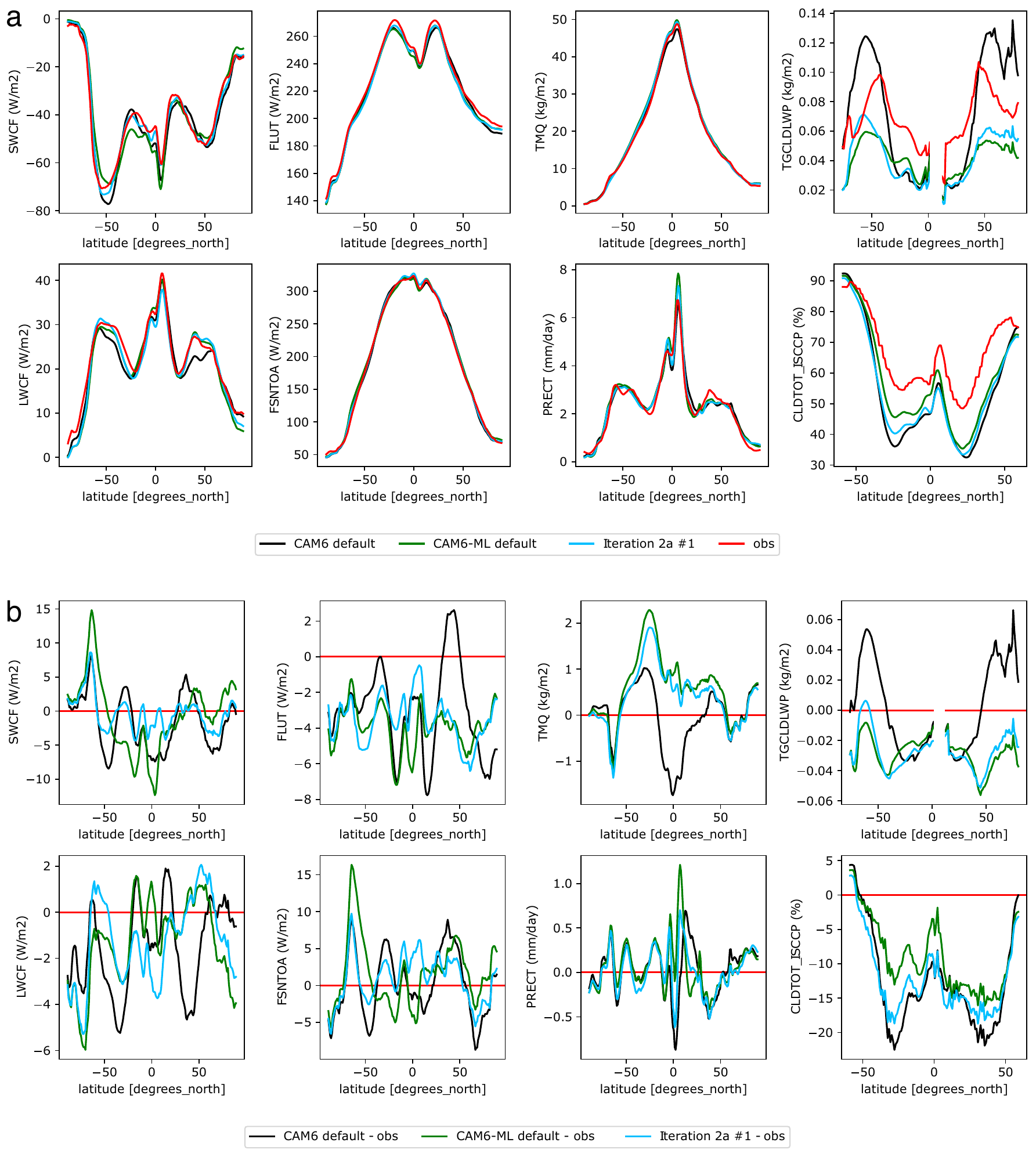}
    \caption{a: Zonal climatologies of the CAM6 default run, the CAM6-ML default run, and the best ensemble member (\#1) from Iteration 2a; b: their differences with satellite observations. }
    \label{f_noodle_diff}
\end{figure}

The best members from Iteration 2a outperform the CAM6 default run in terms of RMSE, but they do not necessarily have smaller global biases. This behavior is illustrated in Fig. \ref{f_noodle_diff}, which compares the zonal climatologies of the Iteration 2a best member ($\#1$), the CAM6 default run, the CAM6-ML default run, and the observations.
The zonal (simulation–observation) differences illustrate why the Iteration 2a best member ($\#1$) does not always have a smaller global bias despite its lower RMSE. This is particularly evident for FLUT (Fig. \ref{f_noodle_diff}). Both the CAM6-ML default run and the Iteration 2a best member display relatively uniform, low-magnitude negative biases across latitudes. In contrast, although the CAM6 default run exhibits larger zonal deviations from the observations (i.e., greater RMSE), it also has a latitude band of positive bias near $\sim 40^\circ$N that partially compensates for the negative biases elsewhere. This spatial compensation reduces its global mean bias, even though the spatial errors are larger in magnitude.

Appendix C provides the zonal climatologies of the best ensemble members from Iterations 2b and 2c, and compares them with observations and the CAM6 default run (CAM6-ML default run omitted for simplicity).

\paragraph{Comparison with the CAM6-ML default run}

It is important to compare the difference between the CAM6-ML default run and the best ensemble members, which could highlight the benefits or limitations using the present method. As mentioned earlier, we are able to find ensemble member (\#1) from Iteration 2a that outperforms both the CAM6-ML default run and the CAM6 default run in terms of RMSE with the exception of CLDTOT$\_$ISCCP.

However, it is hard to make the comparison based on the biases due to the irreducible structural error between LWCF and FLUT. To address this dilemma, we compute the sum of the absolute global biases of the top of the atmosphere radiative flux variables (i.e., SWCF, LWCF, FSNTOA, FLUT) and sum them together (analogous to the $L1$ norm). We refer to this sum as the Aggregated Absolute Bias (AAB), and use this sum as one of the metrics to compare the model performance (Table \ref{t_rmse}). 
We note that this metric has limited physical meaning, and is only proposed for testing the method performance in this work.

All six best ensemble members from Iteration 2a, 2b, and 2c have their AABs lower than that of the CAM6-ML default run (Table \ref{t_rmse}). They are 70-87$\%$ of the magnitude of the AAB of the CAM6-ML default run. Their biases in TMQ are 30-73$\%$ of the CAM6-ML default run. The PRECT biases for the six ensemble members are not uniformly smaller than that of the default run. Those characterized by greater PRECT biases (i.e., Iteration 2b best ensemble members, and Iteration 2c \#71; 153-169$\%$ of the CAM6-ML default run bias) have corresponding smaller biases in FLUT and FSNTOA. The sum of these two absolute biases ranges from 0.57-1.21 W/m$^2$, significantly smaller than that of the CAM6-ML default run (5.41 W/m$^2$). The best ensemble members with greater PRECT biases are characterized by the implicit structural error between the three variables noted earlier. The PRECT biases for the remaining ensemble members are 16-30$\%$ of the CAM6-ML default run bias. This comes at the cost of worsening FSNTOA or FLUT simulation.

The comparison above suggests that parameters selected from our method successfully reduce the AAB relative to the CAM6-ML default run. Without considering the irreducible structural error between LWCF and FLUT, the best ensemble members from Iteration 2a, 2b, and 2c reduce the other biases in two ways: (1) greatly improved FLUT and FSNTOA at the cost of degraded PRECT (Iteration 2b best ensembles and Iteration 2c ensemble \#71); and (2) greatly reduced PRECT with worse FSNTOA or FLUT (the other ensemble members listed in Table \ref{t_rmse}). These two distinct patterns are the result of implicit structural error between PRECT, FSNTOA, and FLUT. The comparison confirms that parameters selected from the method are capable of reducing the biases relative to the CAM6-ML default run.

\subsubsection{The constrained parameter space}
\begin{figure}[H]
   \noindent\includegraphics[width=\textwidth]{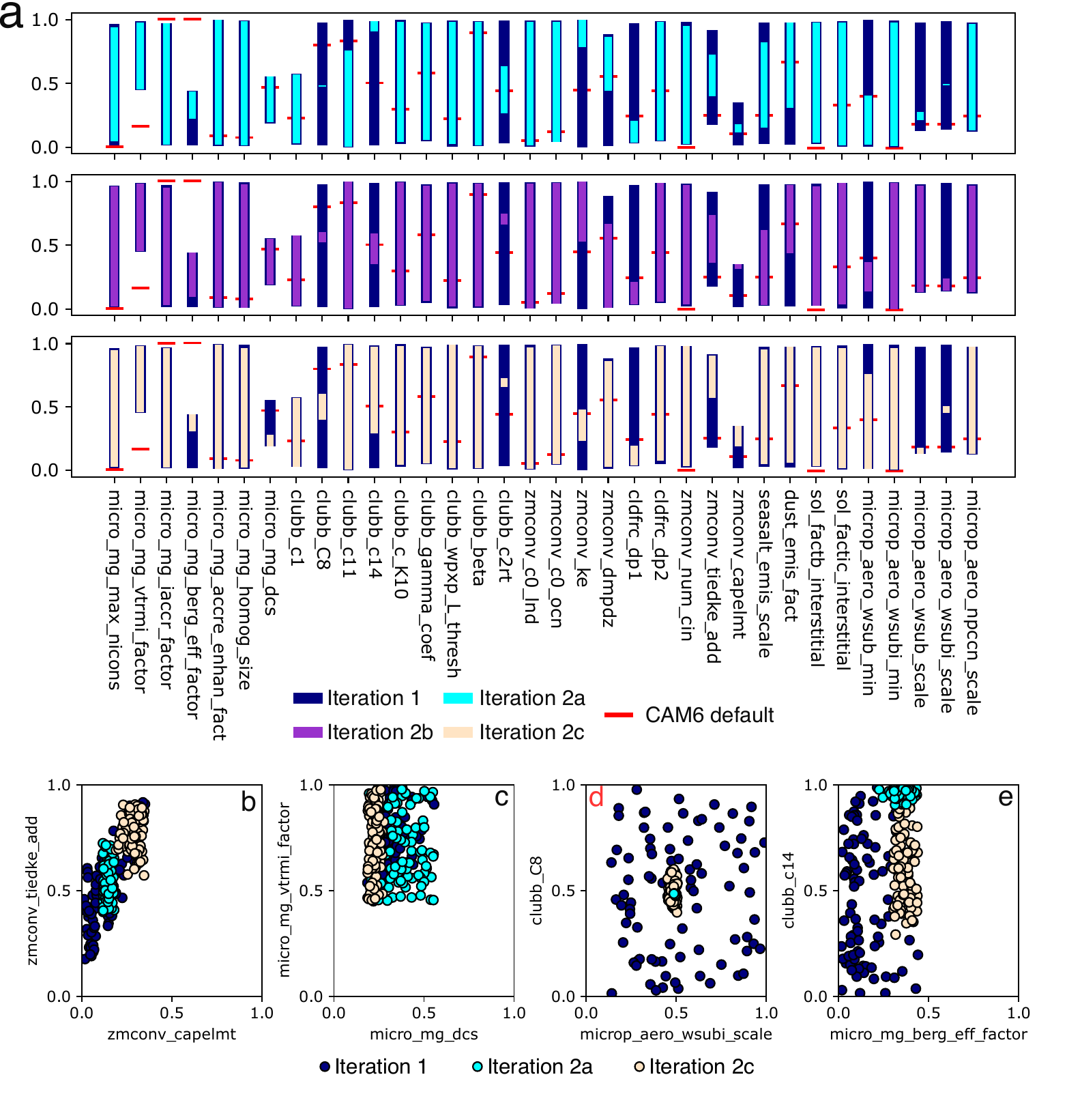}
    \caption{a: Constrained parameter ranges (normalized) from Iterations 1, 2a, 2b, and 2c; b-e: Biplots of selected parameter pairs from Iterations 1, 2a, and 2c. Iteration 2b samples are not shown for simplicity. }
    \label{f_shrink_cor}
\end{figure}

The parameters are greatly constrained in Iterations 2a-2c (Fig. \ref{f_shrink_cor}a). The constrained parameter ranges from Iterations 2a and 2b are different for parameters \textit{clubb$\_$c14}, \textit{clubb$\_$c2rt}, \textit{zmconv$\_$capelmt}, \textit{microp$\_$aero$\_$wsub$\_$scale}, and \textit{microp$\_$aero$\_$wsubi$\_$scale}. The constrained parameter ranges for Iterations 2a and 2c are more consistent with each other. The exceptions are \textit{micro$\_$mg$\_$dcs}, \textit{clubb$\_$c2rt}, \textit{zmconv$\_$ke}, \textit{zmconv$\_$capelmt}.

Some constrained parameters in Iterations 2a-2c lie within very narrow ranges in Fig. \ref{f_shrink_cor}, for example, \textit{microp$\_$aero$\_$wsubi$\_$scale}. Here we use this parameter as an example to evaluate and examine whether over-constraint is happening, and if so, why. This also demonstrates the method's ability to diagnose and interpret its own results.

\begin{figure}[H]
   \noindent\includegraphics[width=\textwidth]{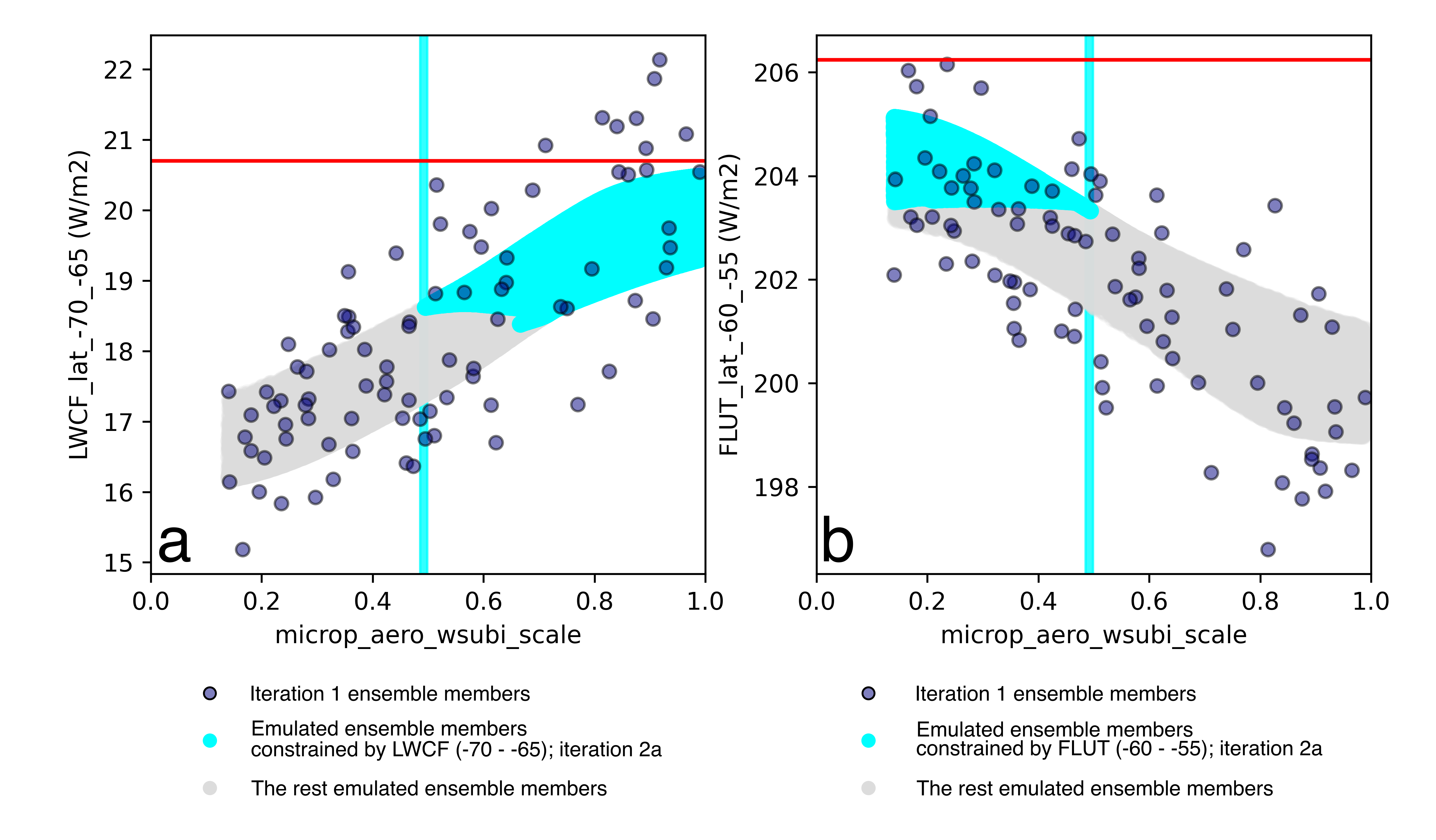}
    \caption{Example showing why $microp\_aero\_wsubi\_scale$ is constrained to a very narrow range in Iteration 2a (Fig. \ref{f_shrink_cor}a). The ensemble members from Iteration 1, and the emulated ensemble member means for targets LWCF from $-70^\circ$S to $-65^\circ$S (a) and FLUT from $-60^\circ$S to $-55^\circ$S (b) are plotted against their most sensitive parameter $microp\_aero\_wsubi\_scale$. The red lines in a and b correspond to the observed values. The cyan vertical strip corresponds to the constrained range of this parameter in Iteration 2a; specifically, the overlapping range of the light blue points. }
    \label{f_detail}
\end{figure}

We first trace back to the target variables that parameter \textit{microp$\_$aero$\_$wsubi$\_$scale} is sensitive to. This is based on the Iteration 1 ensemble as it is used as input to estimate the parameters of Iterations 2a-2c. We find two variables that constrain this parameter the most. They are LWCF from $-70^\circ$S to $-65^\circ$S and FLUT from $-60^\circ$S to $-55^\circ$S. Their values from the Iteration 1 ensemble are plotted against the parameter in Fig. \ref{f_detail}. It shows that the value of \textit{microp$\_$aero$\_$wsubi$\_$scale} needs to be larger to better match the local LWCF and smaller to better match the local FLUT. The cyan points in Fig. \ref{f_detail}a and b are the ensemble members that satisfy $I^k(\mathbf{x}_i^k)=1$ for the two variables at the same time. They have a narrow overlapping range ($I^k(\mathbf{x}_i^k)=1$ for both variables), which is marked by the narrow cyan vertical stripe. As a result, this intrinsic structural error is not considered as a detectable structural error by our method. In the context of this method and its assumptions, our analysis indicates that the parameter \textit{microp$\_$aero$\_$wsubi$\_$scale} is not over-constrained. This example illustrates the interpretability of our proposed method that is capable of explicitly linking how a parameter is constrained to the corresponding target variables.

\begin{figure}[H]
   \noindent\includegraphics[width=\textwidth]{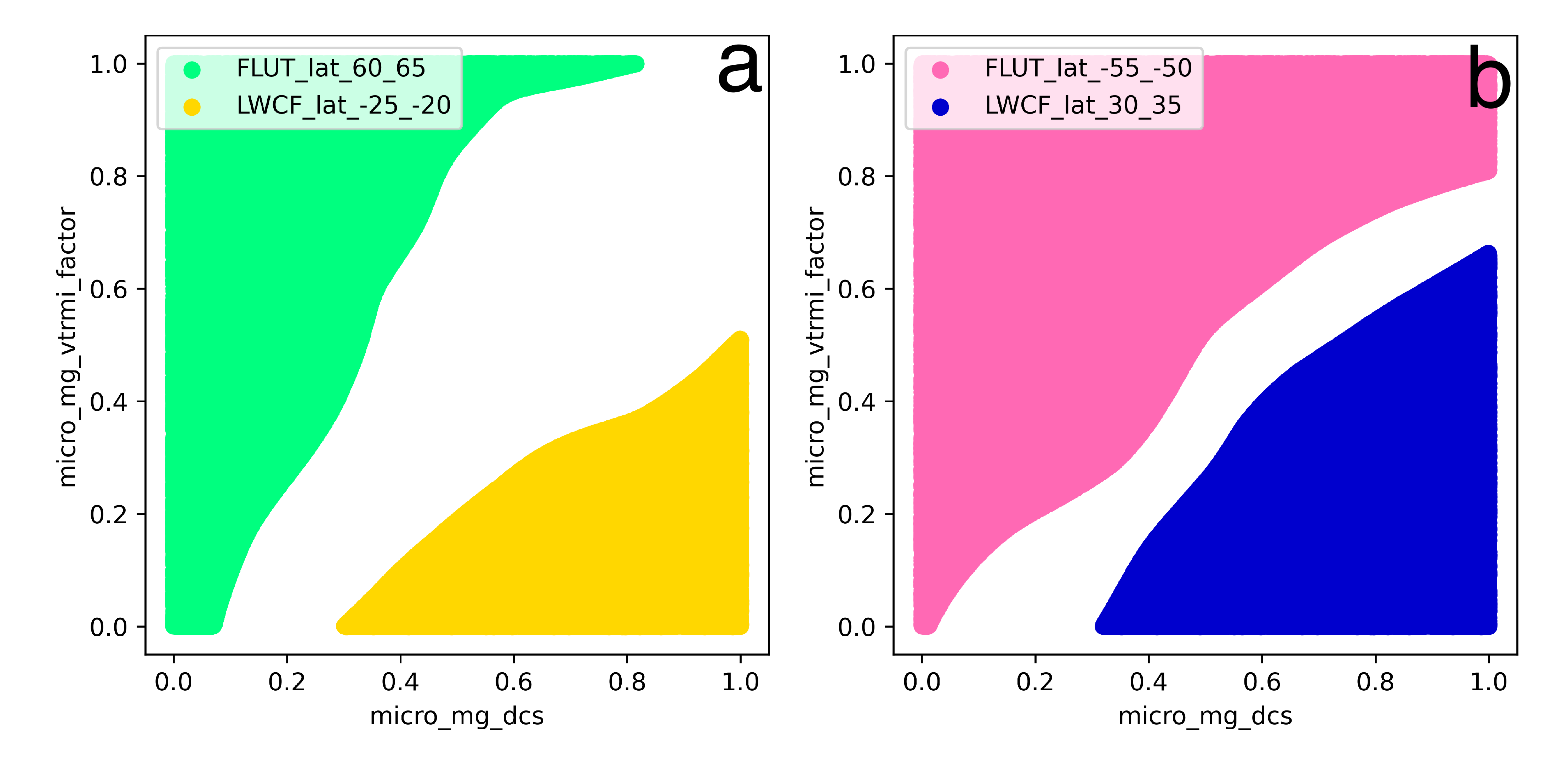}
    \caption{Examples showing the samples satisfying $I^k(\mathbf{x}_i^k) = 1$ for four different target variables based on the Iteration 2a ensemble. The presence of the non-overlapping areas suggests strong structural error. Its common occurrence prevents us from applying the method for another iteration. }
    \label{f_whentostop}
\end{figure}

When we attempt to apply the proposed method to the Iteration 2a ensemble to further constrain the parameters following the CPE methodology, we find many strong structural error between local LWCF and FLUT, and \textit{micro$\_$mg$\_$dcs} and \textit{micro$\_$mg$\_$vtrmi$\_$factor} are their most sensitive parameter pair (Fig. \ref{f_whentostop}). This suggests that we can only improve LWCF at the cost of FLUT or vice versa. For the purpose of this work, we do not proceed with another iteration.

\section{Discussion}
The performance of the proposed method is demonstrated through (1) the reduced spread of the zonal climatologies and their overall convergence towards the observations, (2) the overall reduced global biases and RMSEs, (3) the occurrence of ensemble members that outperform the CAM6 default run in terms of RMSE, (4) the improved model performance compared to the CAM6-ML default run, (5) the narrow ranges of the resultant parameters that are not yet over-constrained (demonstrated by the example of \textit{microp$\_$aero$\_$wsubi$\_$scale}), (6) and the agreement between expectations and results across the different experiments.

\subsection{The impacts of increasing $\tau$}
In Iteration 2c, due to the increased $\tau$, there are fewer local variables to help constrain parameters (Fig. \ref{f_latband}), for example, FSNTOA from $-50^\circ$S to $60^\circ$N. In other words, useful observations are wasted in Iteration 2c. As mentioned earlier, the increased $\tau$ can be viewed as an increase in the assumed emulator uncertainty and/or the introduction of tolerance on structural error. (It can also be viewed as increased observational uncertainty, but this is not considered here as the impact of observational uncertainty is outside the scope of this work.)
Under the interpretation of tolerating the structural error, this waste of information is an intuitive consequence due to the across-the-board increase in $\tau$. In terms of the emulator uncertainty perspective, while the general history-matching literature has recognized this risk under the concept of ``uninformative outputs'' \cite{andrianakis2015bayesian,iskauskas2024emulation}, this consideration has received relatively limited attention within climate model PPE calibration where studies typically report emulator uncertainty validation without connecting it to whether the resulting implausibility threshold leaves individual diagnostics informative.

On the other hand, viewing the increase of $\tau$ together with the decrease in $\phi$ as increasing the tolerance of the structural error, we allow variables that are more likely to bear structural error to be used for parameter estimation. Due to the specific design of the method, this may force estimated parameters to compensate for structural error instead of fitting observations for variables that are free of or less affected by structural error. This is illustrated in the degraded performance of simulated LWCF spanning $\sim30-50^\circ$N (Fig. \ref{f_zonal_bands}) in Iteration 2c. As noted earlier, instead of poorly-constrained LWCF in that latitude range, the Iteration 2c envelope is more constrained and farther away from the observation than Iterations 2a and 2b.

The experimental design of Iteration 2c suggests that its result can be affected by two factors: (1) the uniformly increased $\tau$ would waste useful observations, which leads to poorly constrained parameters. Consequently, this should make the zonal envelope wider instead of narrower than the ones in other iterations. Therefore, the incorrectly constrained LWCF spanning $\sim30-50^\circ$N can only be explained by the second factor, namely (2) the introduction of the structural error tolerance.

\begin{figure}[H]
   \noindent\includegraphics[width=\textwidth]{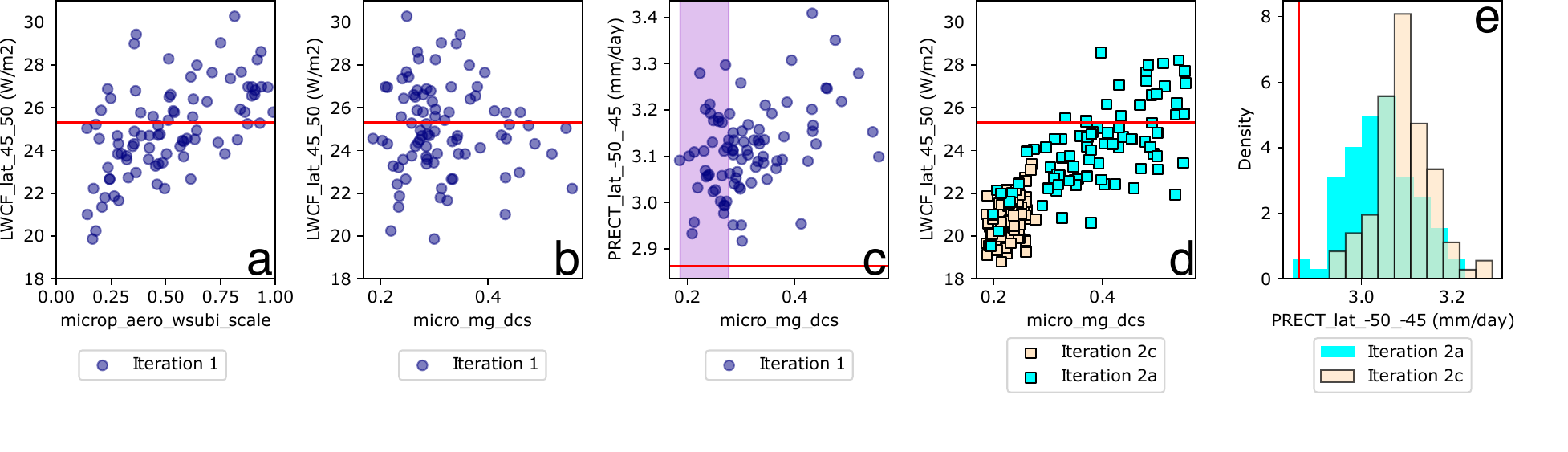}
    \caption{a-c: Variables of interest plotted against parameters \textit{microp$\_$aero$\_$wsubi$\_$scale} and \textit{micro$\_$mg$\_$dcs} from the Iteration 1 ensemble members. In c, the purple vertical stripe shows the parameter range that is constrained by PRECT from $-50^\circ$S to $-45^\circ$S alone in Iteration 2c; d: variable of interest plotted against its most sensitive parameter \textit{micro$\_$mg$\_$dcs} for the Iteration 2a and 2c ensemble members; e: histogram of  PRECT from $-50^\circ$S to $-45^\circ$S from Iterations 2a and 2c ensemble members. Red lines denote the observations in a-e.}
    \label{f_it2c_troubleshoot}
\end{figure}

To understand results further, note that LWCF at $\sim50^\circ$N is sensitive to $micro\_mg\_dcs$ in Iterations 2a and 2c ensembles (Fig. \ref{f_it2c_troubleshoot}d). We notice that the sampled values of $micro\_mg\_dcs$ have a narrower range in Iteration 2c compared to the other iterations (Fig. \ref{f_it2c_troubleshoot}c). By leveraging the interpretability of our method, we find that there is only one variable, namely PRECT from $-50^\circ$S to $-45^\circ$S, that constrains $micro\_mg\_dcs$ to the range in Iteration 2c (Fig. \ref{f_it2c_troubleshoot}c). In Iterations 2a and 2b, this variable is not considered due to the fact that its corresponding $\sum^n_i I^k(\mathbf{x}_i^k)$ is relatively low (which implies that it is likely characterized by strong structural error). It is included in Iteration 2c due to the decreased $\phi$ and increased $\tau$. The observation of PRECT from $-50^\circ$S to $-45^\circ$S constrains $micro\_mg\_dcs$ to an incorrect and smaller range, leading to the degraded performance of LWCF $\sim50^\circ$N in Iteration 2c. This analysis illustrates how tolerating structural error in this method can force the calibration to compensate for model deficiencies, leading to well-constrained yet misleading or incorrect parameters.

The analysis, however, cannot be used to argue that tolerating structural error always degrades the parameter estimation results when other methods are used. This is because our method does not allow assigning weights to different variables as is more common in Bayesian methods, although we could improve the method to enable variable-dependent $\tau$ which could effectively serve as variable weights in addition to differentially weighting particular variables (and their observations); such enhancements and subsequent systematic analyzes would be valuable.

\subsection{Relative parameter importance}
The above example also shows that the parameter sensitivity or importance changes with the constrained region in the parameter space in different iterations of applying the method. We find that LWCF at $\sim50^\circ$N is sensitive to $micro\_mg\_dcs$ based on the Iteration 2a, 2b, and 2c ensembles (Fig. \ref{f_it2c_troubleshoot}d). The same variable is not sensitive to $micro\_mg\_dcs$ but $micro\_aero\_wsubi\_scale$ based on the Iteration 1 ensemble (Fig. \ref{f_it2c_troubleshoot}a and b).
It shows that $micro\_mg\_dcs$ becomes important to LWCF at $\sim50^\circ$N after the other parameters are better constrained, consistent with the findings in \citeA{elsaesser2025using}.

\subsection{Limitations}

\subsubsection{Impacts of intrinsic structural error}
We have shown the negative impacts of tolerating the structural error when this method is used in Iteration 2c. Although our method does not explicitly tolerate the detectable structural error, the method could accidentally or implicitly tolerate the intrinsic structural error. This is because the impact of intrinsic structural error and the emulator predictive uncertainty cannot be completely disentangled (i.e., a value that is considered possible by the emulator predictive uncertainty can be in fact unattainable by the climate model). This suggests that the error arising from the incorrectly constrained \textit{micro$\_$mg$\_$dcs} in Iteration 2c could also occur with use of our method even when the detectable structural error is not tolerated.

\subsubsection{The risk of over-constraining}
The method is designed to be efficient in constraining the parameters at the risk of over-constraining. This is because it is based on history matching, works with many target variables, and views the target variables equally important. This represents a tradeoff between efficiency and over-constraining the parameters in designing the parameter estimation methods. In the case of this method, due to the risk of over-constraining, the method is equipped with the ability to diagnose and interpret its own results as shown in the case with \textit{micro$\_$mg$\_$dcs} in Iteration 2c.

\subsubsection{Over-simplified emulator design}
The method is designed to use only two parameters as inputs to train each of its emulators. The design is tailored to the sparse nature of climate model PPEs, where the limited number of ensemble members creates a risk of overfitting. While the method simplifies the relationships between model parameters and target variables, this trade-off is justified by the curse of dimensionality and the small number of samples that are common in PPE datasets.

\subsubsection{Importance of method interpretability}
The errors associated with the limitations listed above have occurred in our experiments, but the proposed method allows us to trace and identify their causes. This highlights the importance of method interpretability, enabling us to diagnose how parameters are constrained and make informed decisions when irreducible structural errors are present in climate model parameter estimation.

\subsubsection{The possibility of disconnected plausible regions}
The method is able to restrict the parameters into disconnected regions (e.g., a bimodal distribution) in the parameter space. However, if parameters from such different regions are to be used to construct emulators for the next iteration of the simulation–emulation–calibration cycle, users may need to treat the different regions separately to ensure that the training parameters are more evenly distributed within each region.

\section{Conclusions}
In this work, we propose and demonstrate the value of a structural error-aware parameter estimation workflow leveraging climate model Perturbed Parameter Ensembles (PPEs). The method utilizes a history matching approach. Method design choices specifically take into account common characteristics of climate model PPEs, including sparsity in high-dimensional parameter spaces and the presence of structural error and emulator and observational uncertainty.

A key concept proposed in this work is the notion of detectable structural error, defined as systematic mismatches between model outputs and observations that cannot be reconciled through varying the parameter values even after accounting for emulator predictive uncertainty. 
By decomposing the calibration problem into a sequence of low-dimensional subproblems and explicitly detecting and excluding target variables exhibiting detectable structural error, the proposed workflow constrains the parameters using only mutually consistent observations. This design also improves the interpretability of the method, allowing users to directly diagnose which processes or variables are responsible for irreducible, detectable structural error, and detect which variables are responsible for overly constrained parameters. Rather than obscuring structural errors with highly-aggregated misfit metrics or spatial or temporal averages, the workflow makes them explicit at the level of individual targets. In addition, the low-dimensional decomposition enables transparent visualization of constrained regions in the parameter space, providing physical insight into parameter interactions and trade-offs. The proposed approach therefore supports both robust calibration and systematic diagnosis of model error in climate models.

We apply this method to a 100-member PPE dataset generated by CAM6 with a Machine Learning-based warm rain microphysics parameterization. After two iterative applications of the method following the CPE framework \cite{elsaesser2025using}, the resultant ensemble members are greatly constrained towards the observations. We are able to find ensemble members that outperform the CAM6 default run in terms of RMSE. Detectable structural error is identified and explicitly separated from parametric uncertainty for the parameter estimation.

Additional experiments with specific goals are implemented, and the results are consistent with their corresponding hypotheses. Results overall demonstrate the negative impacts of overly conservative emulator uncertainty assumptions and tolerating the structural error when the method is used: (1) the inadvertent discarding of useful observational information that can be otherwise used to constrain the parameters and (2) the inferred parameters being forced to compensate for the strongest structural errors, rather than being constrained toward regions that better fit variables free of (or characterized by less severe) structural error. The second point begs the question of if and how tolerating the structural error would impact the parameter estimation results when other methods are used for climate model PPEs, which requires a more systematic comparative study between history matching- and Bayesian-based methods.

The method has several limitations: (1) since the actual, intrinsic structural error and emulator predictive uncertainty cannot be completely disentangled, the former may still be implicitly tolerated and lead to incorrectly estimated parameters; (2) the method is highly efficient in constraining the parameters, but this efficiency comes with the risk of over-constraining; (3) the use of low-dimensional emulators, while determined to be necessary under sparse PPE sampling, may oversimplify joint parameter effects and lead to incorrectly constrained parameters when three or more parameters all strongly influence a target. The method can yield disconnected constrained regions in the parameter space. When such regions arise, they should be treated separately in subsequent emulator training and parameter estimation to ensure balanced sampling and robust inference. These limitations highlight the importance of the interpretability of the method, which allows users to diagnose how the parameters are constrained and apply domain expertise to make informed decisions throughout the calibration process.

Our results further demonstrate that parameter sensitivity evolves as the constrained region in the parameter space shrinks. Parameters that appear unimportant in the initial ensemble may become important after other parameters are constrained, highlighting the iterative and state-dependent nature of PPE-based calibration.

The proposed workflow provides a practical and interpretable framework for parameter estimation in complex Earth system models in the presence of irreducible structural error. By combining history matching, low-dimensional decomposition and integration, and explicit structural error detection, the method enables efficient exploration of a high-dimensional parameter space while maintaining transparency in the inference process.

\section*{Figures}

\section*{Tables}
\begin{table}[H]
    \caption{Example showing how the method automatically excludes target variables that share the same sensitive parameter pair and display structural errors. The four variable pairs share the same sensitive parameters (micro$\_$mg$\_$berg$\_$eff$\_$factor and microp$\_$aero$\_$wsubi$\_$scale), and are considered to display structural error (the criterion or threshold is determined by the user in a similar way to $\phi$). LWCF from -70$^\circ$S to -65$^\circ$ occurs most frequently in this table, and will be excluded by the method in the first place. This table as one output of the method also informs manual selection of structural error-prone variables if needed.}
    \noindent\includegraphics[width=1.5\textwidth]{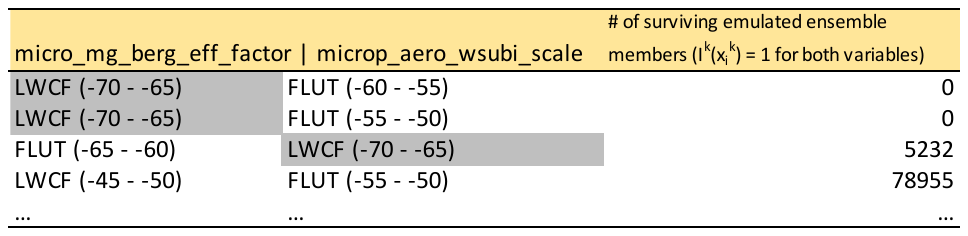}
    \label{t_strt_err}
\end{table}

\begin{table}[H]
    \caption{Parameter names, descriptions, ranges that are used to generate the original 100 member PPE. The CAM6 default values are also given. }
    \noindent\includegraphics[width=\textwidth]{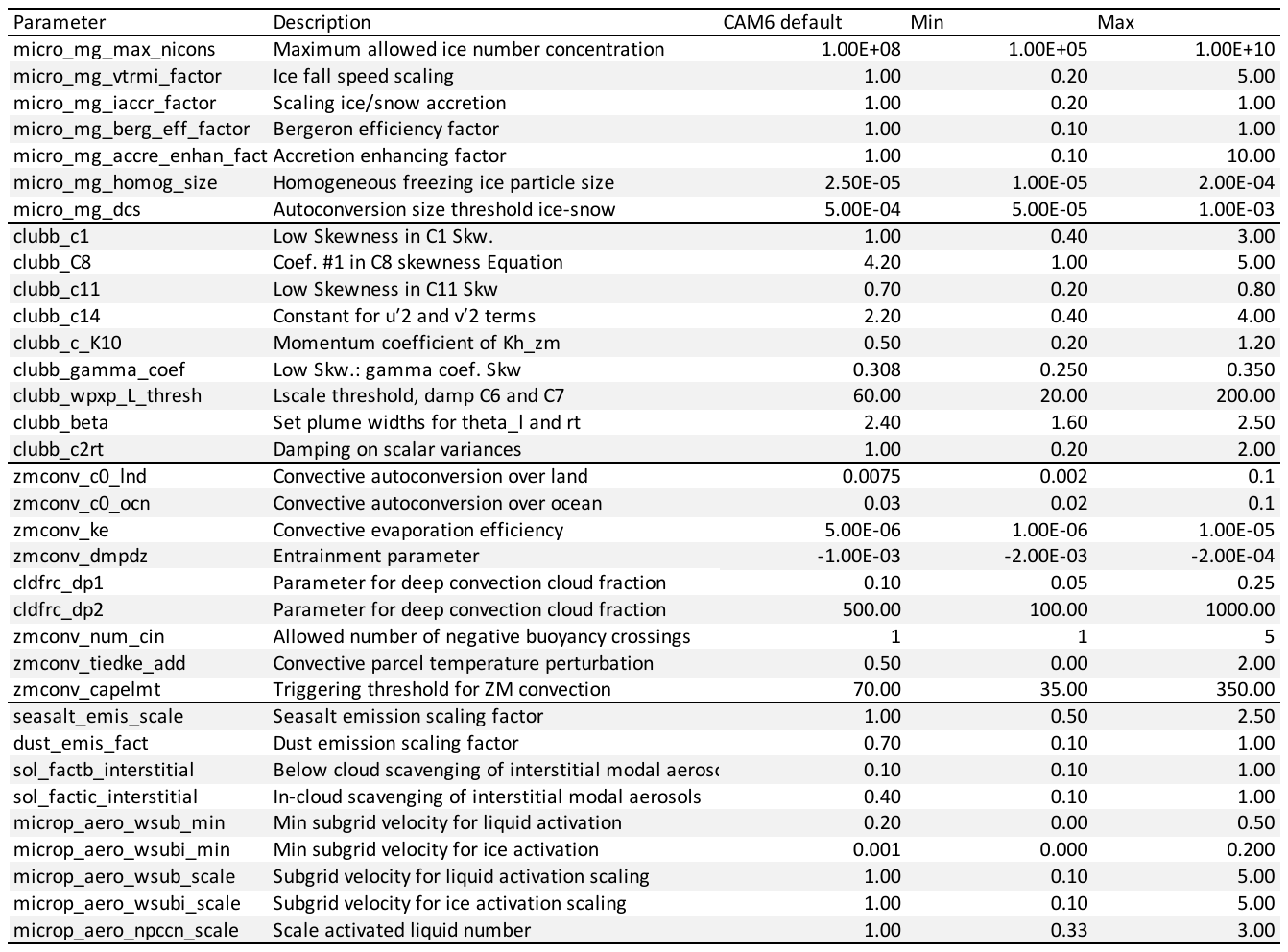}
    \label{t_para_descrip}
\end{table}

\begin{table}[H]
    \caption{Observational products used in this work.}
    \noindent\includegraphics[width=\textwidth]{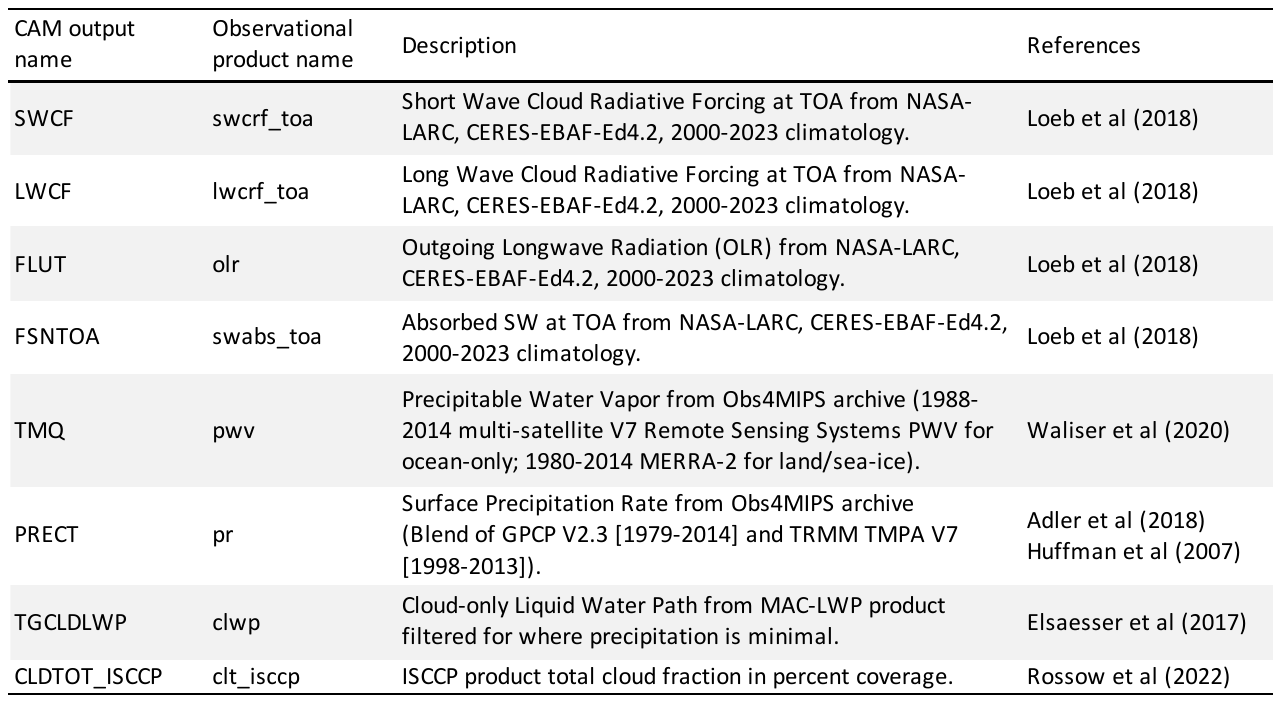}
    \label{t_obs_descrip}
\end{table}

\begin{table}[H]
    \caption{Global bias and RMSE of the best two ensemble members from Iterations 1, 2a, 2b, and 2c. Cells marked in blue are those that outperform the CAM6 default run. We do not mark the cells for TGCLDLWP and CLDTOT$\_$ISCCP because they are not the main concerns due to strong structural error.}
    \noindent\includegraphics[width=\textwidth]
    {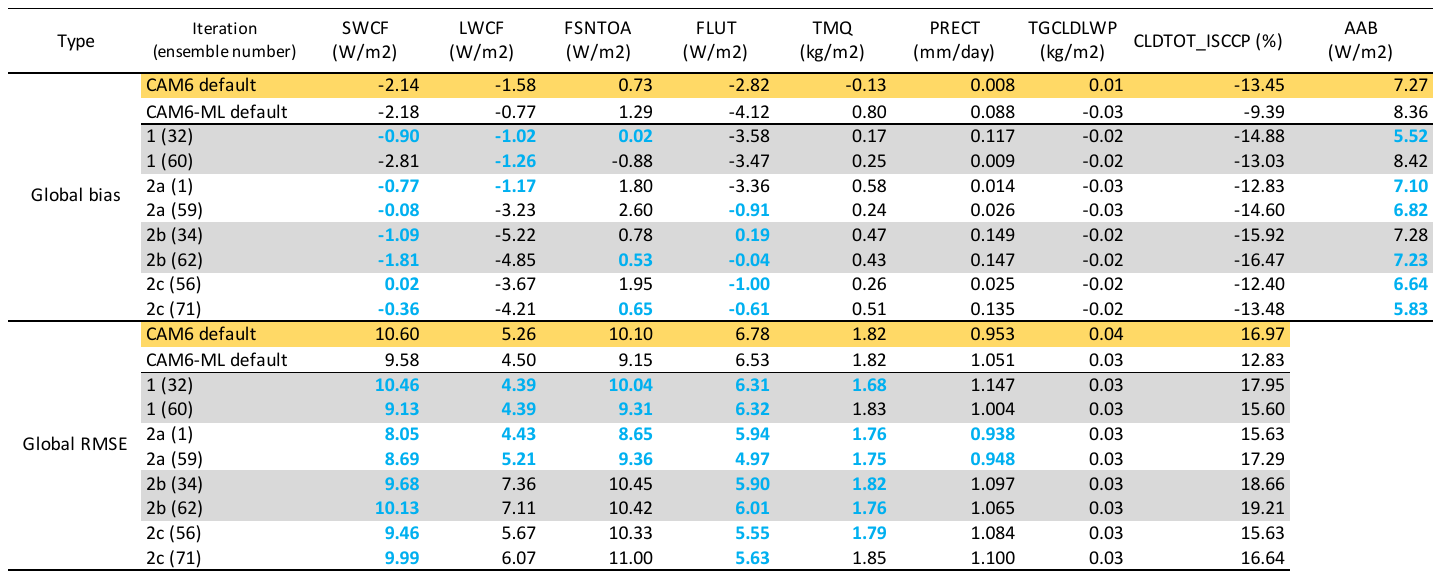}
    \label{t_rmse}
\end{table}

\newpage

\section*{Appendix A}
Selected model output in addition to the sampled parameters.
\begin{figure}[H]
\includegraphics[width=\textwidth]{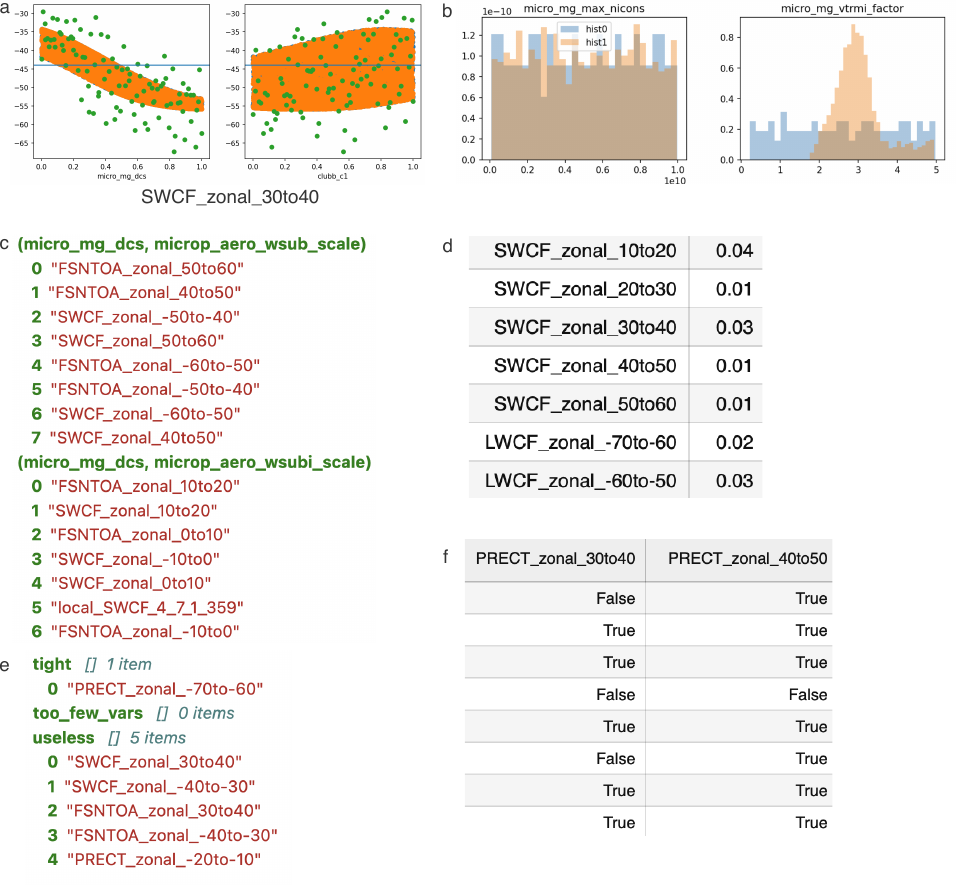}
\caption*{Selected model output in addition to the sampled parameters. Subsets of the outputs are shown here due to the limit on page size. a: Selected target variables plotted against the two most sensitive parameters used for emulation. The green and orange points correspond to the simulated (PPE or CPE) and emulated ensemble members; b: histogram of PPE or CPE parameters and that of the sampled parameters from the method; c: how different target variables are grouped by their most sensitive parameters; d: emulator error rate for each target variables (the ratio of PPE or CPE members that are outside the uncertainty defined by $\tau$ and the predicted GP uncertainty);  e: variables that are excluded for parameter estimation; f: the value of $I^k(\mathbf{x}_i^k)$ for each variable and for each potential sample parameters. }
\end{figure}

\newpage
\section*{Appendix B}
Test results obtained by randomly drawing one PPE member as the observation and using the remaining members as inputs to the method. Ten tests were conducted in total. The volume ratio reported in the table is the product of the sampled ranges (max-min) across all parameters, divided by the corresponding product of parameter ranges in the original PPE. The true parameter values all lie within the ranges of the estimated parameter samples from using the method. 
\begin{table}[H]
    \centering
    \caption*{Test results from Leave-One-Out test.}
    \includegraphics[width=\textwidth]{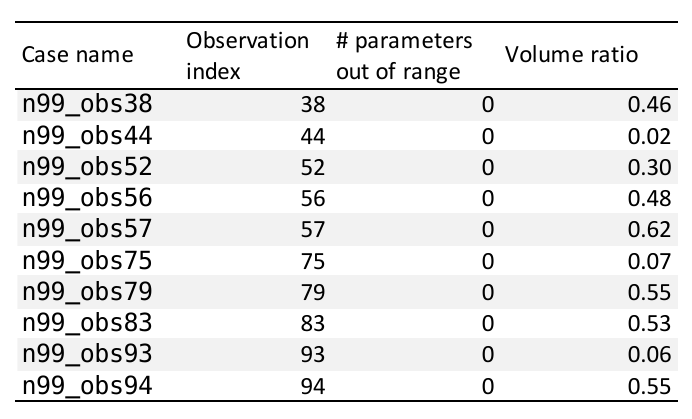}
\end{table}

\newpage
\section*{Appendix C}

\begin{figure}[H]
   \noindent\includegraphics[width=\textwidth]{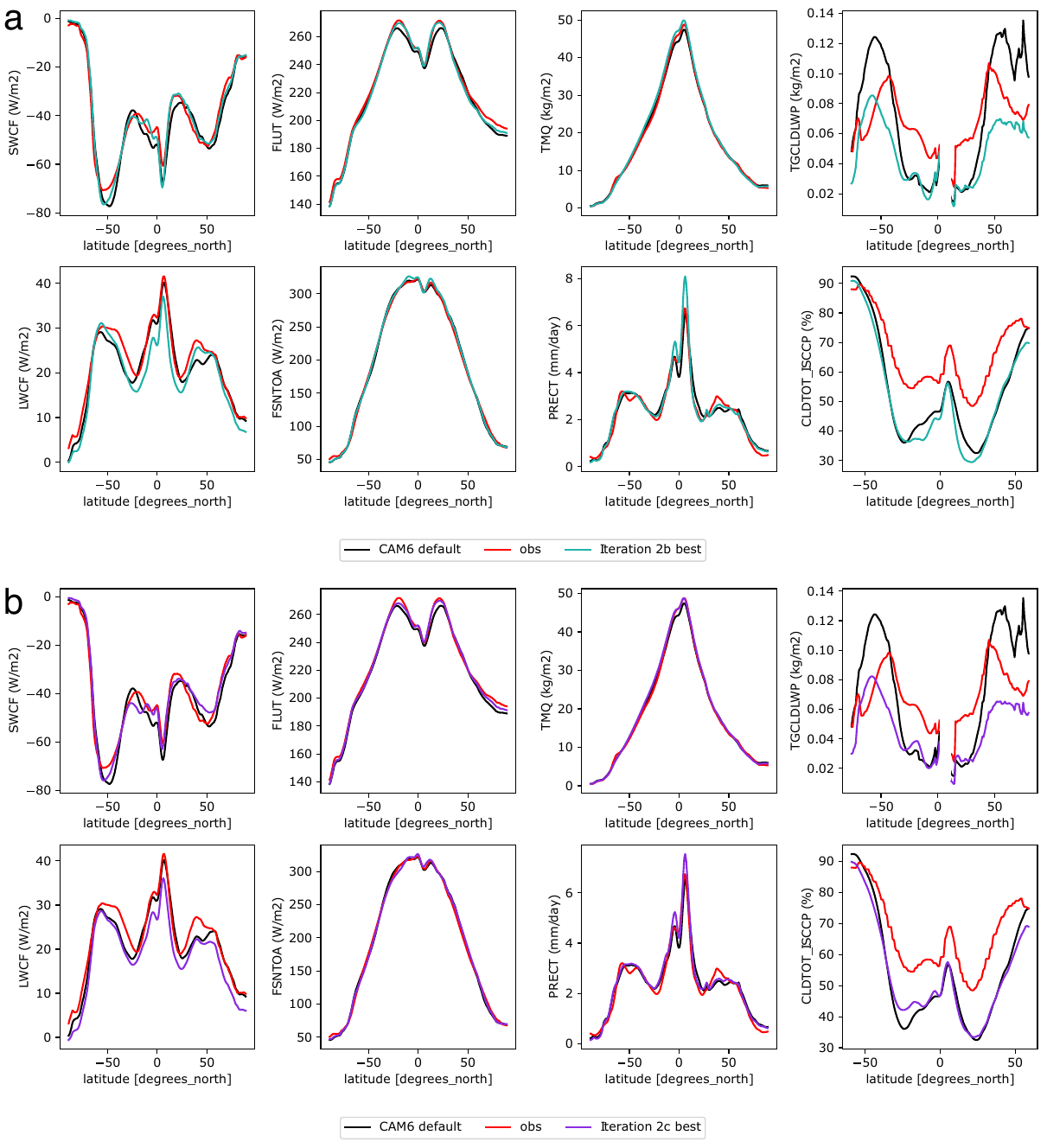}
    \caption*{a-c: Zonal climatologies of the best ensemble members from Iteration 2b and 2c respectively. Observations and CAM6 default run are plotted for reference. }
    \label{app_noodle}
\end{figure}

\section*{Open Research Statement}
The datasets used in this work are made available on Zenodo (\url{https://doi.org/10.5281/zenodo.21904973}). Upon acceptance of the manuscript, the repository will be officially published and a permanent DOI will be provided. The method is available on github (\url{https://github.com/yiqioyang/proj2dhullsampler}).

\acknowledgments
 We acknowledge funding from NSF through the Learning the Earth with Artificial intelligence and Physics (LEAP) NSF Science and Technology Center (STC) (Award \#2019625).  We acknowledge additional support from NASA-MAP (including grants \#80NSSC21K1498 and \#80NSSC17K1499). We thank  L. Hawkins and P. Gentine for the insightful discussions on PPE analysis.

\bibliography{main}

\end{document}